%% file: main.tex
\pdfoutput=1
\documentclass[12pt,a4paper]{article}
\usepackage{ifthen} 
\usepackage{multirow}
\newboolean{pdflatex}
\setboolean{pdflatex}{true} 

\newboolean{articletitles}
\setboolean{articletitles}{true} 

\newboolean{uprightparticles}
\setboolean{uprightparticles}{false} 

\newboolean{inbibliography}
\setboolean{inbibliography}{false} 

\def\paperauthors{LHCb collaboration} 
\def\paperasciititle{Measurement of Upsilon production in pp collisions at sqrt{s}=13 TeV} 
\def\papertitle{Measurement of~$\ups$~production in $pp$~collisions at~$\sqs=13\tev$} 
\def\paperkeywords{{High Energy Physics}, {LHCb}} 
\def\papercopyright{\the\year\ CERN for the benefit of the LHCb collaboration} 
\def\paperlicence{CC-BY-4.0 licence}
\def\paperlicenceurl{https://creativecommons.org/licenses/by/4.0/}

\RequirePackage[normalem]{ulem} 
\RequirePackage{color}\definecolor{RED}{rgb}{1,0,0}\definecolor{BLUE}{rgb}{0,0,1} 

\input{preamble}
\usepackage{longtable} 
\usepackage{rotating}
\begin{document}

\renewcommand{\thefootnote}{\fnsymbol{footnote}}
\setcounter{footnote}{1}

\input{title-LHCb-PAPER}

\renewcommand{\thefootnote}{\arabic{footnote}}
\setcounter{footnote}{0}

\pagestyle{plain} 
\setcounter{page}{1}
\pagenumbering{arabic}

\input{up-paper-body}

\input{acknowledgements}

\addcontentsline{toc}{section}{References}
\setboolean{inbibliography}{true}
\bibliographystyle{LHCb}
\bibliography{main.bbl}

\newpage

\newpage
\input{LHCb_Authorship_flat_22-Jan-2018.tex}

\end{document}

%% file: preamble.tex
\usepackage[top=1in, bottom=1.25in, left=1in, right=1in]{geometry}

\usepackage{microtype}
\usepackage{lineno}  
\usepackage{xspace} 
\usepackage{caption} 

\usepackage{graphicx}  
\usepackage{color}
\usepackage{colortbl}
\graphicspath{{./figs/}} 

\usepackage{amsmath} 
\usepackage{amssymb}
\usepackage{amsfonts}
\usepackage{upgreek} 

\newcommand*\patchAmsMathEnvironmentForLineno[1]{%
\expandafter\let\csname old#1\expandafter\endcsname\csname #1\endcsname
\expandafter\let\csname oldend#1\expandafter\endcsname\csname
end#1\endcsname
 \renewenvironment{#1}%
   {\linenomath\csname old#1\endcsname}%
   {\csname oldend#1\endcsname\endlinenomath}%
}
\newcommand*\patchBothAmsMathEnvironmentsForLineno[1]{%
  \patchAmsMathEnvironmentForLineno{#1}%
  \patchAmsMathEnvironmentForLineno{#1*}%
}
\AtBeginDocument{%
\patchBothAmsMathEnvironmentsForLineno{equation}%
\patchBothAmsMathEnvironmentsForLineno{align}%
\patchBothAmsMathEnvironmentsForLineno{flalign}%
\patchBothAmsMathEnvironmentsForLineno{alignat}%
\patchBothAmsMathEnvironmentsForLineno{gather}%
\patchBothAmsMathEnvironmentsForLineno{multline}%
\patchBothAmsMathEnvironmentsForLineno{eqnarray}%
}

\usepackage{hyperxmp}

\usepackage[pdftex,
            pdfauthor={\paperauthors},
            pdftitle={\paperasciititle},
            pdfkeywords={\paperkeywords},
            pdfcopyright={Copyright (C) \papercopyright},
            pdflicenseurl={\paperlicenceurl}]{hyperref}

\usepackage[all]{hypcap} 

\input{lhcb-symbols-def} 

\usepackage{cite} 
\usepackage{mciteplus}

%% file: lhcb-symbols-def.tex
	\def\lhcb {\mbox{LHCb}\xspace}

	\def\MagUp {\mbox{\em Mag\kern -0.05em Up}\xspace}

	\ifthenelse{\boolean{uprightparticles}}%
{

	\def\Pmu         {\ensuremath{\upmu}\xspace}

	\def\Ppsi        {\ensuremath{\uppsi}\xspace}

	\def\PDelta      {\ensuremath{\Delta}\xspace}                 
	\def\PXi      {\ensuremath{\Xi}\xspace}                 
	\def\PLambda      {\ensuremath{\Lambda}\xspace}                 
	\def\PSigma      {\ensuremath{\Sigma}\xspace}                 
	\def\POmega      {\ensuremath{\Omega}\xspace}                 
	\def\PUpsilon      {\ensuremath{\Upsilon}\xspace}                 

	\def\PB      {\ensuremath{\mathrm{B}}\xspace}                 
	                 
	\def\PD      {\ensuremath{\mathrm{D}}\xspace}

	\def\PJ      {\ensuremath{\mathrm{J}}\xspace}                 
	\def\PK      {\ensuremath{\mathrm{K}}\xspace}

	\def\Pb      {\ensuremath{\mathrm{b}}\xspace}                 
	\def\Pc      {\ensuremath{\mathrm{c}}\xspace}

	\def\Pi      {\ensuremath{\mathrm{i}}\xspace}

}
{

	\def\Pmu         {\ensuremath{\mu}\xspace}

	\def\Ppsi        {\ensuremath{\psi}\xspace}                 
	                 
	\mathchardef\PDelta="7101
		\mathchardef\PXi="7104
		\mathchardef\PLambda="7103
		\mathchardef\PSigma="7106
		\mathchardef\POmega="710A
		\mathchardef\PUpsilon="7107
		                 
	\def\PB      {\ensuremath{B}\xspace}                 
	                 
	\def\PD      {\ensuremath{D}\xspace}

	\def\PJ      {\ensuremath{J}\xspace}                 
	\def\PK      {\ensuremath{K}\xspace}

	\def\Pb      {\ensuremath{b}\xspace}                 
	\def\Pc      {\ensuremath{c}\xspace}

	\def\Pi      {\ensuremath{i}\xspace}

}

\makeatletter
\ifcase \@ptsize \relax
\newcommand{\miniscule}{\@setfontsize\miniscule{4}{5}}
\or
\newcommand{\miniscule}{\@setfontsize\miniscule{5}{6}}
\or
\newcommand{\miniscule}{\@setfontsize\miniscule{5}{6}}
\fi
\makeatother

\DeclareRobustCommand{\optbar}[1]{\shortstack{{\miniscule (\rule[.5ex]{1.25em}{.18mm})}
	\\ [-.7ex] $#1$}}

	\def\mumu       {{\ensuremath{\Pmu^+\Pmu^-}}\xspace}

	\def\cquark    {{\ensuremath{\Pc}}\xspace}

	\def\bquark    {{\ensuremath{\Pb}}\xspace}

	\def\Kbar    {{\kern 0.2em\overline{\kern -0.2em \PK}{}}\xspace}
	
	\def\KorKbar    {\kern 0.18em\optbar{\kern -0.18em K}{}\xspace}

	\def\Dbar    {{\kern 0.2em\overline{\kern -0.2em \PD}{}}\xspace}

	\def\DorDbar    {\kern 0.18em\optbar{\kern -0.18em D}{}\xspace}

	\def\Bbar    {{\ensuremath{\kern 0.18em\overline{\kern -0.18em \PB}{}}}\xspace}
	
	\def\BorBbar    {\kern 0.18em\optbar{\kern -0.18em B}{}\xspace}

	\def\jpsi     {{\ensuremath{{\PJ\mskip -3mu/\mskip -2mu\Ppsi\mskip 2mu}}}\xspace}

	\def\Lbar        {{\ensuremath{\kern 0.1em\overline{\kern -0.1em\PLambda}}}\xspace}
	\def\LorLbar    {\kern 0.18em\optbar{\kern -0.18em \PLambda}{}\xspace}

	\def\BF         {{\ensuremath{\cal B}}\xspace}
	
	\def\BR         {\BF}
	
	\def\to                 {\ensuremath{\rightarrow}\xspace}

  \newcommand{\ups}    {\PUpsilon}

  \newcommand{\OneS}  {\ensuremath{\ups{(1S)}}\xspace}
  \newcommand{\TwoS}  {\ensuremath{\ups{(2S)}}\xspace}
  \newcommand{\ThreeS}{\ensuremath{\ups{(3S)}}\xspace}

	\newcommand{\etot}{{\ensuremath{\varepsilon_{\rm tot}}}\xspace}

	\def\AT#1     {\ensuremath{A_{\mathrm{T}}^{#1}}\xspace}           

	\def\C#1      {\ensuremath{\mathcal{C}_{#1}}\xspace}                       
	\def\Cp#1     {\ensuremath{\mathcal{C}_{#1}^{'}}\xspace}                    
	\def\Ceff#1   {\ensuremath{\mathcal{C}_{#1}^{\mathrm{(eff)}}}\xspace}        
	\def\Cpeff#1  {\ensuremath{\mathcal{C}_{#1}^{'\mathrm{(eff)}}}\xspace}       
	\def\Ope#1    {\ensuremath{\mathcal{O}_{#1}}\xspace}                       
	\def\Opep#1   {\ensuremath{\mathcal{O}_{#1}^{'}}\xspace}                    

	\newcommand{\tev}{\ifthenelse{\boolean{inbibliography}}{\ensuremath{~T\kern -0.05em eV}\xspace}{\ensuremath{\mathrm{\,Te\kern -0.1em V}}}\xspace}
	\newcommand{\gev}{\ensuremath{\mathrm{\,Ge\kern -0.1em V}}\xspace}
	\newcommand{\mev}{\ensuremath{\mathrm{\,Me\kern -0.1em V}}\xspace}
	\newcommand{\kev}{\ensuremath{\mathrm{\,ke\kern -0.1em V}}\xspace}
	\newcommand{\ev}{\ensuremath{\mathrm{\,e\kern -0.1em V}}\xspace}
	\newcommand{\gevc}{\ensuremath{{\mathrm{\,Ge\kern -0.1em V\!/}c}}\xspace}
	\newcommand{\mevc}{\ensuremath{{\mathrm{\,Me\kern -0.1em V\!/}c}}\xspace}
	\newcommand{\gevcc}{\ensuremath{{\mathrm{\,Ge\kern -0.1em V\!/}c^2}}\xspace}
	\newcommand{\gevgevcccc}{\ensuremath{{\mathrm{\,Ge\kern -0.1em V^2\!/}c^4}}\xspace}
	\newcommand{\gevgevcc}{\ensuremath{{\mathrm{\,Ge\kern -0.1em V^2\!/}c^2}}\xspace}
	\newcommand{\mevcc}{\ensuremath{{\mathrm{\,Me\kern -0.1em V\!/}c^2}}\xspace}

	\def\mum  {\ensuremath{{\,\upmu\rm m}}\xspace}

	\def\pb {\ensuremath{\rm \,pb}\xspace}
	\def\invpb {\ensuremath{\mbox{\,pb}^{-1}}\xspace}

	\def\deriv {\ensuremath{\mathrm{d}}}

	\def\gsim{{~\raise.15em\hbox{$>$}\kern-.85em
		\lower.35em\hbox{$\sim$}~}\xspace}
		\def\lsim{{~\raise.15em\hbox{$<$}\kern-.85em
			\lower.35em\hbox{$\sim$}~}\xspace}

			\def\sqs   {\ensuremath{\protect\sqrt{s}}\xspace}

			\def\ptot       {\mbox{$p$}\xspace}
			\def\pt         {\mbox{$p_{\rm T}$}\xspace}

			\newcommand{\lum} {\ensuremath{\mathcal{L}}\xspace}

			\def\evtgen     {\mbox{\textsc{EvtGen}}\xspace}

			\def\geant      {\mbox{\textsc{Geant4}}\xspace}

			\def\photos     {\mbox{\textsc{Photos}}\xspace}

			\def\pythia     {\mbox{\textsc{Pythia}}\xspace}

			\def\tell1  {TELL1\xspace}
			\def\ukl1   {UKL1\xspace}



%% file: title-LHCb-PAPER.tex
\begin{titlepage}
\pagenumbering{roman}

\vspace*{-1.5cm}
\centerline{\large EUROPEAN ORGANIZATION FOR NUCLEAR RESEARCH (CERN)}
\vspace*{1.5cm}
\noindent
\begin{tabular*}{\linewidth}{lc@{\extracolsep{\fill}}r@{\extracolsep{0pt}}}
\ifthenelse{\boolean{pdflatex}}
{\vspace*{-1.5cm}\mbox{\!\!\!\includegraphics[width=.14\textwidth]{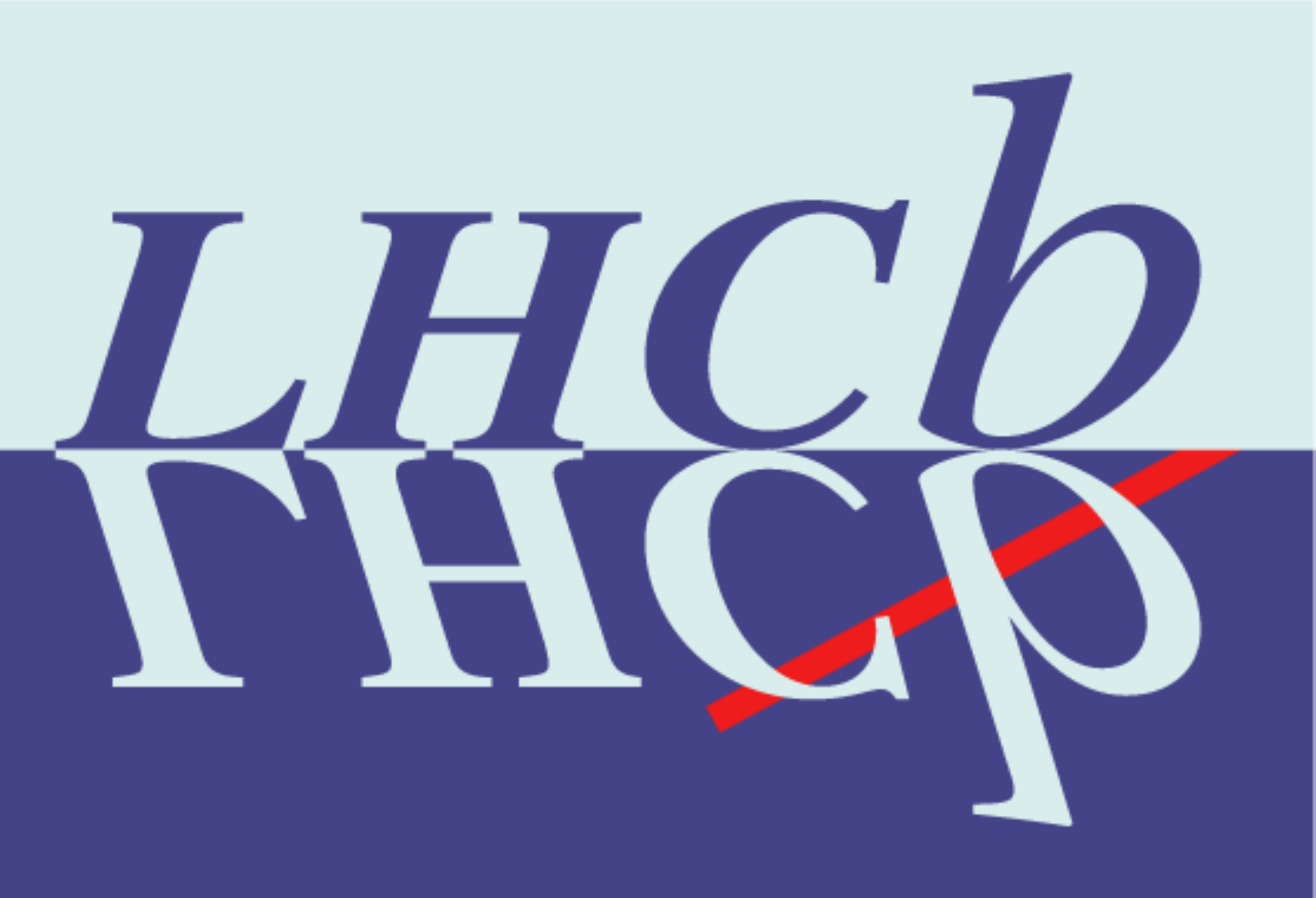}} & &}%
{\vspace*{-1.2cm}\mbox{\!\!\!\includegraphics[width=.12\textwidth]{lhcb-logo.eps}} & &}%
\\
 & & CERN-EP-2018-054 \\  
 & & LHCb-PAPER-2018-002 \\  
 & & 1 August 2018 \\ 
 & & \\
\end{tabular*}

\vspace*{4.0cm}

{\normalfont\bfseries\boldmath\huge
\begin{center}
  \papertitle 
\end{center}
}

\vspace*{2.0cm}

\begin{center}
\paperauthors\footnote{Authors are listed at the end of this paper.}
\end{center}

\vspace{\fill}

\begin{abstract}
  \noindent
   The production cross-sections of \OneS, \TwoS and \ThreeS
  mesons in proton-proton collisions at $\sqrt{s}=13\tev$ are measured 
  with a data sample corresponding to an integrated luminosity of $277\pm11\invpb$ recorded by the LHCb experiment in 2015.
  The $\ups$ mesons are reconstructed in the decay mode
	$\ups\to\mu^{+}\mu^{-}$.  
	The differential production cross-sections times the 
	dimuon branching
  fractions are measured as a function of the $\ups$ transverse
  momentum, $\pt$, and rapidity, $y$, over the range 
  $0<\pt<30\gevc$ and $2.0<y<4.5$. 
  The ratios of the cross-sections with respect to the LHCb
  measurement at $\sqrt{s}=8\tev$ are also determined.
  The measurements are compared with theoretical predictions based on
  NRQCD. 
\end{abstract}

\vspace*{2.0cm}

\begin{center}
  Published in JHEP 07(2018)134
\end{center}

\vspace{\fill}

{\footnotesize 
\centerline{\copyright~\papercopyright. \href{\paperlicenceurl}{\paperlicence}.}}
\vspace*{2mm}

\end{titlepage}

\newpage
\setcounter{page}{2}
\mbox{~}

\cleardoublepage

%% file: up-paper-body.tex
\section{Introduction}
\label{sec:Introduction}
The study of heavy quarkonium ($c\overline{c}$ and $b\overline{b}$) production in high-energy hadron collisions provides important 
information to better understand quantum chromodynamics (QCD). 
Thanks to theoretical and experimental efforts in the past forty
years, the comprehension of hadronic production 
of heavy quarkonia has been improved significantly.  
The production of heavy quarkonium in proton-proton ($pp$) collisions at the Large Hadron Collider (LHC)
is expected to start with 
the production of a heavy
quark pair, $Q\overline{Q}$, followed by its
hadronization into a bound state.
The heavy quark pair $Q\overline{Q}$ is produced mainly via Leading
Order (LO) gluon-gluon interactions.
Several models have been proposed to describe 
the underlying dynamics, such as the colour-singlet
model (CSM)~\cite{Carlson:1976cd,Donnachie:1976ue,Ellis:1976fj,Fritzsch:1977ay,Gluck:1977zm,Chang:1979nn,Baier:1981uk}
and non-relativistic QCD (NRQCD)~\cite{Bodwin:1994jh,Cho:1995vh,Cho:1995ce}.
In the CSM the intermediate $Q\overline{Q}$ state is supposed to be colourless 
and has the same quantum numbers as the quarkonium final state, 
while in NRQCD the calculations also include the colour-octet
contribution. 
However, at present no model can describe both 
the heavy quarkonium production cross-section and polarisation simultaneously. 

The production of $\OneS$, $\TwoS$ and $\ThreeS$ mesons has been
studied at LHC 
by ATLAS~\cite{ATLAS:Upsilon,ATLAS:Upsilon7TeV} and
CMS~\cite{CMS:Upsilon7TeV,CMS:quarkonium} collaborations at 
centre-of-mass energies of 
$7\tev$ and $13\tev$.
The measurements have also been performed by LHCb at 
centre-of-mass energies of $2.76\tev$~\cite{LHCb-PAPER-2013-066}, 
$7\tev$~\cite{LHCb-PAPER-2011-036,LHCb-PAPER-2015-011,LHCb-PAPER-2015-045,LHCb-PAPER-2017-028}
and
$8\tev$~\cite{LHCb-PAPER-2013-016, LHCb-PAPER-2015-011,LHCb-PAPER-2015-045,LHCb-PAPER-2017-028}.
The NRQCD calculations can describe the trends of the differential production cross-sections in data for all three $\ups$ states within uncertainties. 
The measured production cross-section ratios between $7\tev$ and
$8\tev$~\cite{LHCb-PAPER-2015-045} as a function of transverse momentum, $\pt$, 
are consistently higher than the next-to-leading order NRQCD theory predictions~\cite{Han:2014kxa},
and the ratios as a function of rapidity show a different trend than the predictions~\cite{Kisslinger:2013mev,Kisslinger:2014zga}.
The measurement performed at $13\tev$ presented here provides 
valuable input to study 
the quarkonium production at a higher centre-of-mass energy, 
enabling ratios to be determined with respect to 
data taken at a lower centre-of-mass energy.
Most of the theoretical and experimental
uncertainties cancel in the ratios, and more stringent
constraints on the theoretical models can be obtained.

This paper presents a measurement of  
the differential production cross-sections times dimuon branching
fractions of 
$\OneS$, $\TwoS$ and $\ThreeS$ mesons,
as functions of $\pt$, and rapidity, $y$, over the range $0<\pt<30\gevc$ and $2.0<y<4.5$. 
Ratios between the cross-section measurements at $13\tev$ and
$8\tev$~\cite{LHCb-PAPER-2015-045} are also presented.
The measurements are compared with theoretical predictions 
based on NRQCD~\cite{Wang:Upsilon2015}.

\section{The \lhcb detector and event selection}
\label{sec:Detector}
The \lhcb detector~\cite{Alves:2008zz,LHCb-DP-2014-002} is a single-arm forward
spectrometer covering the \mbox{pseudorapidity} range $2<\eta <5$,
designed for the study of particles containing \bquark or \cquark
quarks. The detector includes a high-precision tracking system
consisting of a silicon-strip vertex detector surrounding the $pp$
interaction region~\cite{LHCb-DP-2014-001}, a large-area silicon-strip detector located
upstream of a dipole magnet with a bending power of about
$4{\mathrm{\,Tm}}$, and three stations of silicon-strip detectors and straw
drift tubes~\cite{LHCb-DP-2013-003} placed downstream of the magnet.
The tracking system provides a measurement of momentum, \ptot, of charged particles with
a relative uncertainty that varies from 0.5\% at low momentum to 1.0\% at 200\gevc.
The minimum distance of a track to a primary vertex, the impact parameter,
is measured with a resolution of $(15+29/\pt)\mum$, where $\pt$ is expressed in \gevc.
Different types of charged hadrons are distinguished using information
from two ring-imaging Cherenkov detectors~\cite{LHCb-DP-2012-003}.
Photons, electrons and hadrons are identified by a calorimeter system consisting of
scintillating-pad (SPD) and preshower detectors, an electromagnetic
calorimeter and a hadronic calorimeter. Muons are identified by a
system composed of alternating layers of iron and multiwire
proportional chambers~\cite{LHCb-DP-2012-002}.

The data sample used in this measurement corresponds to an integrated luminosity of 
 $277\pm11\invpb$ of $pp$ collisions at a centre-of-mass energy of
 $13\tev$ collected during 2015.
The online event selection is performed by a trigger system~\cite{LHCb-DP-2012-004}
that consists
of a hardware stage selecting 
dimuon candidates with the product of the transverse momenta of the muons
greater than $(1.3 \gev/c)^2$,
followed by a two-stage software selection
based on the information available after 
full event reconstruction.
In the software trigger, two muons
with $p>6\gevc$, $\pt>300\mevc$ are selected to form a $\ups$ candidate.
These two muon candidates are required to form a common vertex~\cite{Hulsbergen:2005pu} with
an invariant mass $M>4.7\gevcc$.
To reject high-multiplicity events with a large number of $pp$
interactions, a set of global event requirements~\cite{LHCb-DP-2012-004} is applied,
which includes the requirement that the number of hits in the
SPD subdetector be less than 900.
In between the first and second stages of the software trigger, 
the alignment and calibration of the detector is performed 
     nearly in real-time \cite{LHCb-PROC-2015-011} and updated constants are made
     available for the
     trigger. The same alignment and calibration information is propagated
     to the offline reconstruction, ensuring 
		 the consistency and high-quality of the tracking and 
     particle identification information between the trigger and
     offline software. The identical performance of the online and offline
     reconstruction offers the opportunity to perform physics analyses
     directly using candidates reconstructed in the trigger
     \cite{LHCb-DP-2012-004,LHCb-DP-2016-001}
     as done in the present analysis.

After the trigger, the $\ups$ candidates are further selected offline by requiring
two well identified muon candidates with 
transverse momentum larger than $1\gevc$, 
and momentum larger than $10\gevc$. 
The invariant mass of the two muon candidates is required to be in the range $8.5 < M < 11.5\gevcc$.
Furthermore, 
the muons are required to form a vertex with good fit quality. 

The event-selection efficiencies are determined using 
simulated samples, which are generated using
\pythia8~\cite{Sjostrand:2007gs} 
with a specific LHCb configuration~\cite{LHCb-PROC-2010-056}.  
The three $\ups$ states are assumed to be produced unpolarised in this
analysis. 
The decays of hadrons are described by \evtgen~\cite{Lange:2001uf}, 
in which final-state radiation is simulated using \photos~\cite{Golonka:2005pn}. 
The \geant toolkit~\cite{Allison:2006ve} is used to describe the interactions of the generated particles with the detector and its response.

\section{Cross-section determination}
\label{sec:cross}
The double-differential production cross-section times dimuon branching
fraction ($\BR$) is defined as
\begin{equation}
  \frac{\deriv^2\sigma}{\deriv y\deriv \pt} \times\BR(\ups\to\mumu)
 = \frac{N(\pt,y)}
         {\lum\times\etot(\pt,y)\times\Delta y \times \Delta \pt}, \label{eq:CrossSec}
\end{equation}
where $N(\pt,y)$ and $\etot(\pt,y)$
are respectively the signal yields and the total efficiencies for 
$\OneS$, $\TwoS$ and $\ThreeS$ states in the given kinematic bin,
$\lum$ is the integrated luminosity, and $\Delta \pt$ and $\Delta y$ are the bin widths. 

To determine the signal yields in each kinematic bin, an extended
unbinned maximum-likelihood fit is performed to the dimuon invariant mass distribution of the selected candidates. 
The dimuon mass distribution is described by three Crystal Ball
functions~\cite{Skwarnicki:1986xj}, one for each of the three $\ups$ states,
and the combinatorial background is described by 
an exponential function. 
In each bin, the tail parameters of the Crystal Ball functions
are fixed as done in the previous analysis~\cite{LHCb-PAPER-2011-036}.
The mass and mass resolution of the $\OneS$ state are free
parameters. For the $\TwoS$ and $\ThreeS$ states, the mass differences
with respect to the $\OneS$ state are fixed to the current world
averages~\cite{PDG2016} and 
the ratios of the mass resolutions with respect to that of the $\OneS$ state
are fixed to those in the simulated samples.

\begin{figure}
\centering
\begin{minipage}[t]{0.6\textwidth}
\centering
\includegraphics[width=1.0\textwidth]{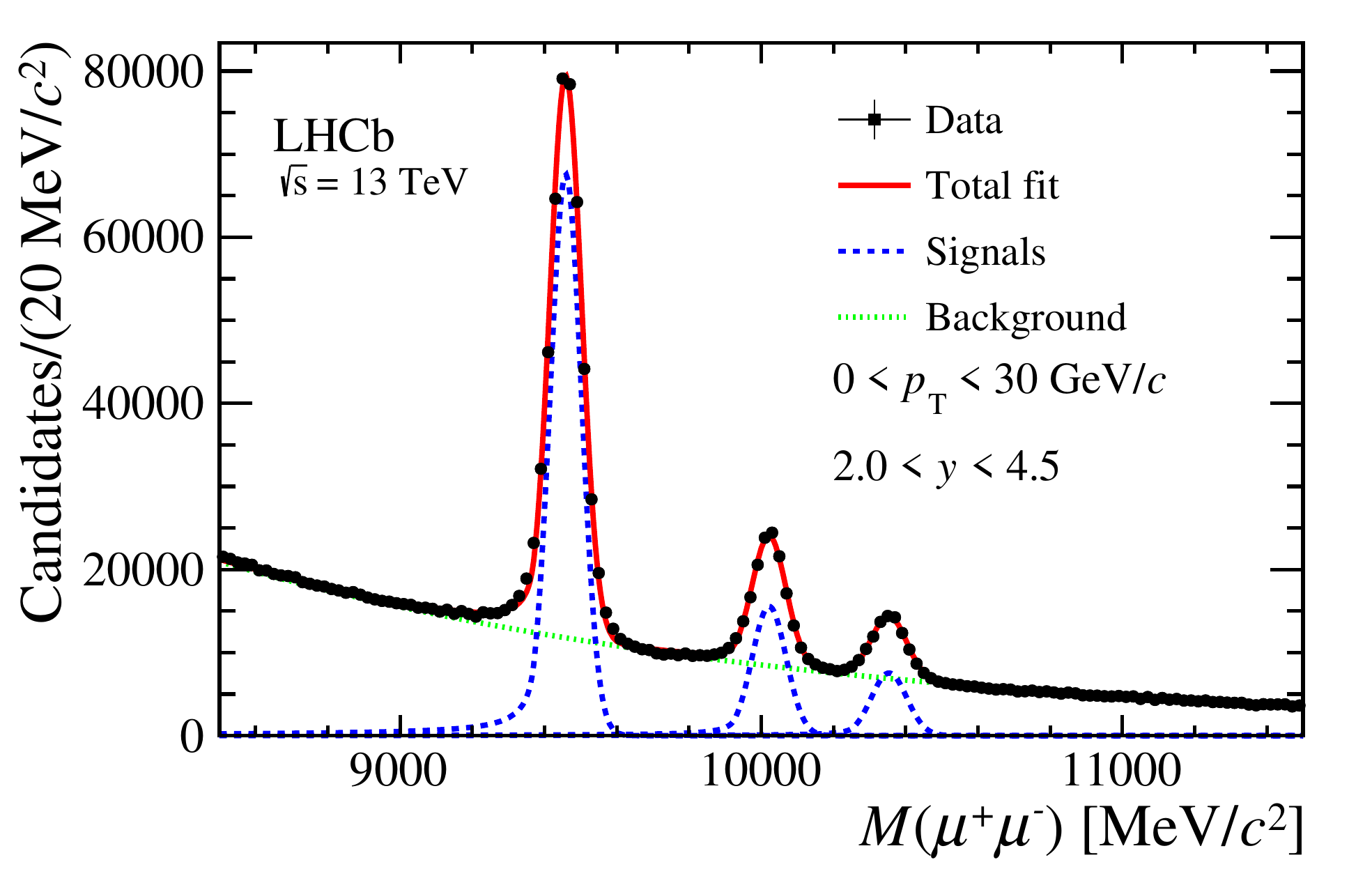}
\end{minipage}
\caption{\small Dimuon invariant mass distribution of $\ups$ candidates
  with $0<\pt<30\gevc$ and $2.0<y<4.5$. The fit result with
  the Crystal Ball functions plus an exponential function is also
  shown. 
  The black dots refer to the data, the blue dashed line refers to the three
  signals and the green dotted line refers to the background. }
\label{fig::MM}
\end{figure}

The invariant mass distribution of $\ups$ candidates after all the selections in the range
$0<\pt<30\gevc$ and $2.0<y<4.5$
and the fit results are shown in Fig.~\ref{fig::MM}.  
The total signal yields are $397\,841\pm796$ for the $\OneS$ state,
$99\,790\pm469$ for the $\TwoS$ state and $50\,677\pm381$ for the $\ThreeS$ state.

The total efficiency, $\etot$, in each bin is computed as the product
of the detector geometrical 
acceptance and of the efficiencies related to particle reconstruction,
event selection, muon identification and trigger. 
The detector acceptance, selection and trigger efficiencies are calculated
using simulation. 
The tracking efficiency is obtained from simulation 
and corrected using a data-driven method~\cite{LHCb-DP-2013-002} to improve its modelling in 
simulation samples. 
The muon identification efficiency is
determined from simulation and calibrated with
$\jpsi\to\mumu$ and $\phi\to\mumu$ data using a tag-and-probe method.
The calibration samples in data are not sufficient to 
give precise tracking and muon identification efficiencies 
for the whole kinematic region. 

\section{Systematic uncertainties}
\label{sec:sys}
Several sources of systematic uncertainties are associated with the determination of the
signal yields, efficiencies, and integrated luminosity. They are 
reported in Table~\ref{tab:SysSum} and described below. The dominant
uncertainties are due to the trigger efficiency and the luminosity.

\begin{table}
\begin{center}  
\begin{tabular}{|p{4.6cm}cccc|}
\hline
Source & \OneS &\TwoS &\ThreeS  & Comment \\
\hline
Fit models & 1.9&1.8 &2.5 & Correlated\\
Simulation statistics & $0.4 - 4.6$& $0.5 - 5.1$& $0.5 - 4.4$& Bin dependent\\
Global event requirements& 0.6&0.6&0.6& Correlated\\
Trigger&$ 3.9 - 9.8$&$3.9 - 9.8$&$3.9 - 9.8$ & Bin dependent\\
\multirow{2}{*}{Tracking}& $(0.1-6.6)$ & $(0.2-6.4)$ & $(0.2-6.5)$  &\multirow{2}{*}{ Correlated}\\
  &$\oplus (2\times0.8)$ & $\oplus (2\times0.8)$& $\oplus (2\times0.8)$   &   \\
Muon identification & $0.1 - 7.9$&$ 0.1 - 7.6$& $0.2 - 8.5$ & Correlated\\
Vertexing& 0.2 & 0.2 & 0.2 & Correlated\\
Kinematic spectrum &$ 0.0 - 1.1$&$ 0.0 - 2.2$&$0.0 - 2.5$ & Bin dependent\\
Radiative tail & 1.0& 1.0 & 1.0 & Correlated\\ 
Luminosity& 3.9& 3.9 & 3.9& Correlated\\
\hline
Total  & $6.2-14.3$ & $6.2-14.6$ & $6.4-14.9$ &Correlated\\
\hline
\end{tabular}
\end{center}
\caption{Summary of the relative systematic uncertainties (in $\%$) on
  the $\ups$ production cross-sections times dimuon branching
  fractions. Some of the uncertainties are correlated between
  intervals. 
For the trigger, track reconstruction and muon identification
efficiencies, the uncertainties are larger in the high rapidity
region.
The uncertainties on the tracking efficiency account for both 
the limited size of the control samples (first parenthesis) and 
the impact of different multiplicity between data and simulation on each track (second parenthesis).
}
\label{tab:SysSum}
\end{table}

The uncertainty related to the fit model predominantly originates from the
description of the signal tails caused by final-state
radiation, and from the description of the background line shape. 
The former is studied 
by using the same fit model to describe the dimuon invariant-mass distribution of  
a mixed sample in which the signal is from the full simulation, and
the background is generated with pseudoexperiments using the same shape and fraction as in the data.
The latter is studied by replacing the  
exponential function with a second-order Chebyshev polynomial function.
A combined relative uncertainty of $1.9\%$ for the $\OneS$ state, $1.8\%$ for the $\TwoS$ state and $2.5\%$ for the $\ThreeS$ state is assigned.

The efficiency of the global event requirements 
is found to be 100\% in simulation, and 99.4\% in data, with
negligible statistical uncertainty. The difference, 0.6\%, is assigned as 
systematic uncertainty. 

The trigger efficiency uncertainty is computed using the
same method as 
in the 7$\tev$ and 8$\tev$ analyses~\cite{LHCb-PAPER-2015-045}. 
The dimuon hardware-trigger efficiency is studied 
with events triggered by the single-muon hardware trigger. 
The difference of this efficiency in data and simulated
samples divided by that in simulation
is assigned as the relative systematic uncertainty. 
The systematic uncertainty related to the global event requirements
applied in the single-muon trigger, 
as well as other sources of uncertainties
are assumed to be equal to those estimated
in the inclusive $b\bar{b}$ cross-section measurements at 7$\tev$ and
13$\tev$~\cite{LHCb-PAPER-2016-031}.
In total, the systematic uncertainty coming from trigger efficiencies is
$3.9-9.8\%$ for the three $\ups$ states, depending on the kinematic bin.

The tracking efficiencies in simulation are 
corrected with a data-driven method using $\jpsi\to\mumu$ control samples. 
The number of SPD hits distributions in simulated samples are weighted
to improve the agreement with data and
an uncertainty of $0.8\%$ per track is assigned to account for
a different multiplicity between data and simulation.
The tracking efficiencies determined from simulated samples are corrected with the ratio of efficiencies in data and simulation. 
The uncertainty
due to the finite size of the
control samples is propagated to the final systematic uncertainty
using a large number of pseudoexperiments
and is $0.1-6.6\%$ for the $\OneS$ state, $0.2-6.4\%$ for the $\TwoS$ state and $0.2-6.5\%$ for the $\ThreeS$
state, depending on the kinematic bin.
In total, the systematic uncertainty originating  
from the tracking efficiency is $1.6-6.8\%$ for the $\OneS$ state,
$1.6-6.6\%$ for the $\TwoS$ state and $1.6-6.7\%$ for the $\ThreeS$
state, depending on the kinematic bin.

The muon identification efficiency is determined from simulation and 
calibrated with data using 
a tag-and-probe method. 
The single-muon identification efficiency is measured 
in intervals of $p$, 
$\eta$ and event multiplicity. 
The statistical uncertainty
due to the finite size of the calibration sample 
is propagated to the final 
results using pseudoexperiments. 
The uncertainty 
related to the kinematic binning scheme of the calibration samples 
is studied by changing the bin size. 
The uncertainty due to the kinematic correlations between
the two muons, which is not considered in the efficiency calculation,
is studied with simulated samples. 
The correlation is found to be negligible except for the most forward
region, in which the two muons from the $\ups$ decays have smaller opening angle.
In total, the systematic uncertainty assigned to the muon
identification efficiency is $0.1-7.9\%$ for the $\OneS$ state, $0.1-7.6\%$ for
the $\TwoS$ state and $0.2-8.5\%$ for the $\ThreeS$ state, depending on the kinematic bin. 

The systematic uncertainty on the signal efficiency of the vertex fit quality requirement is studied by comparing data and
simulation. A relative difference of $0.2\%$ is assigned for the three $\ups$ states.

The kinematic distributions of $\ups$ mesons in simulation and in data
are slightly different within each kinematic bin due to the finite bin
size in $\pt$ and $y$, 
causing differences in efficiencies.
This effect is studied by weighting the kinematic distributions of
$\ups$ states in simulation to match the distributions in data. 
All efficiencies are recalculated, and the relative
differences of the total efficiency between the new and the
nominal results are assigned as systematic uncertainties,
which vary between $0.0-1.1\%$ for the $\OneS$ state, $0.0-2.2\%$ for
the $\TwoS$ state and $0.0-2.5\%$ for the $\ThreeS$ state, depending on the kinematic bin. 

A systematic uncertainty of $1.0\%$ is assigned 
as a consequence of the limited precision on the modelling of the
final-state radiation in the simulation, 
estimated as in the previous analysis~\cite{LHCb-PAPER-2013-016}.

The integrated luminosity is determined using the beam-gas imaging and 
van der Meer scan methods~\cite{LHCb-PAPER-2014-047}. 
A relative uncertainty of $3.9\%$ is assigned on the luminosity 
and propagated to the cross-sections.

\section{Results}
\subsection{Cross-sections}
\begin{figure}[!tbp]
\centering
\begin{minipage}[t]{0.6\textwidth}
\centering
\includegraphics[width=1.0\textwidth]{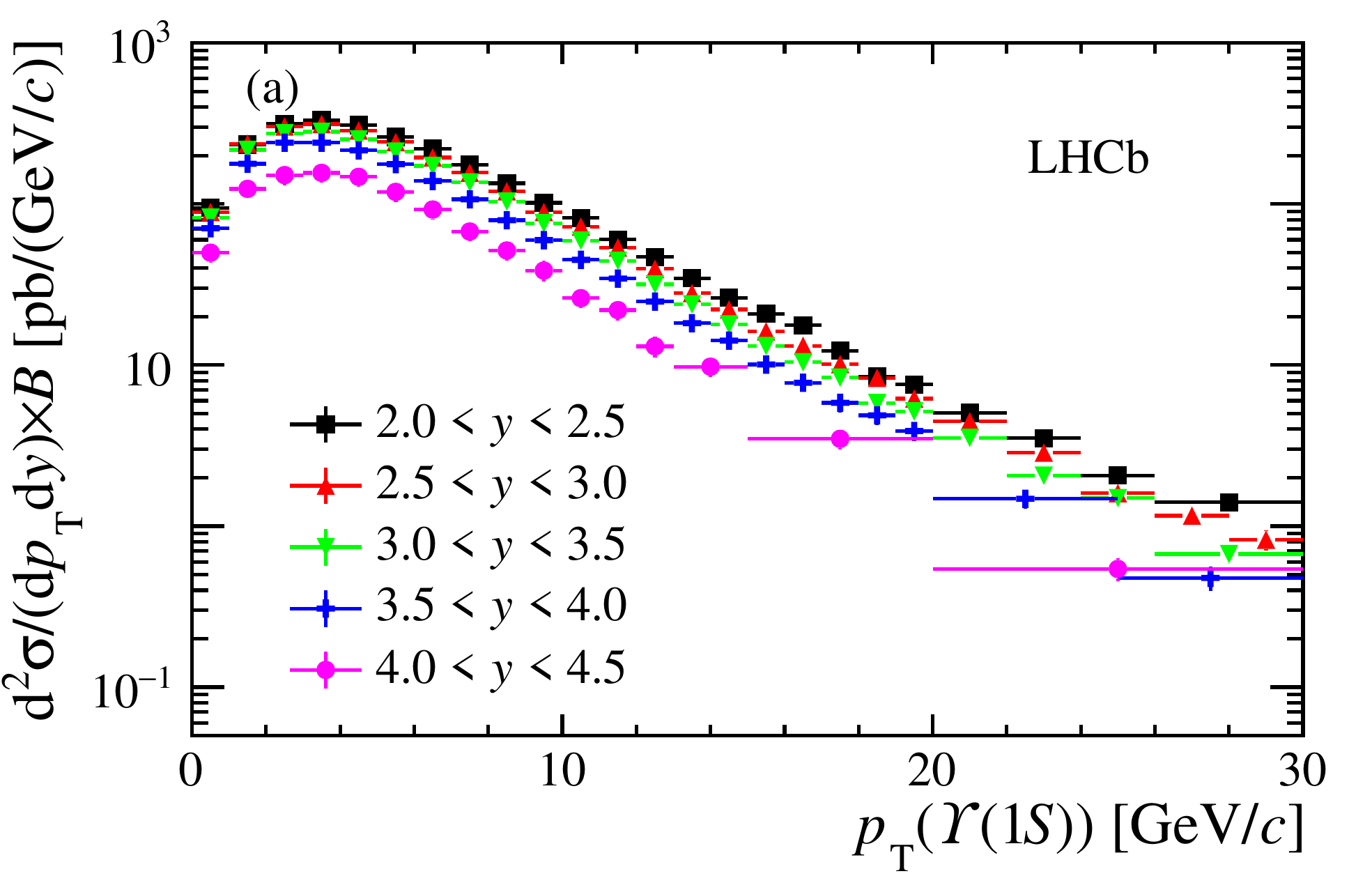}
\end{minipage}
\centering
\begin{minipage}[t]{0.6\textwidth}
\centering 
\includegraphics[width=1.0\textwidth]{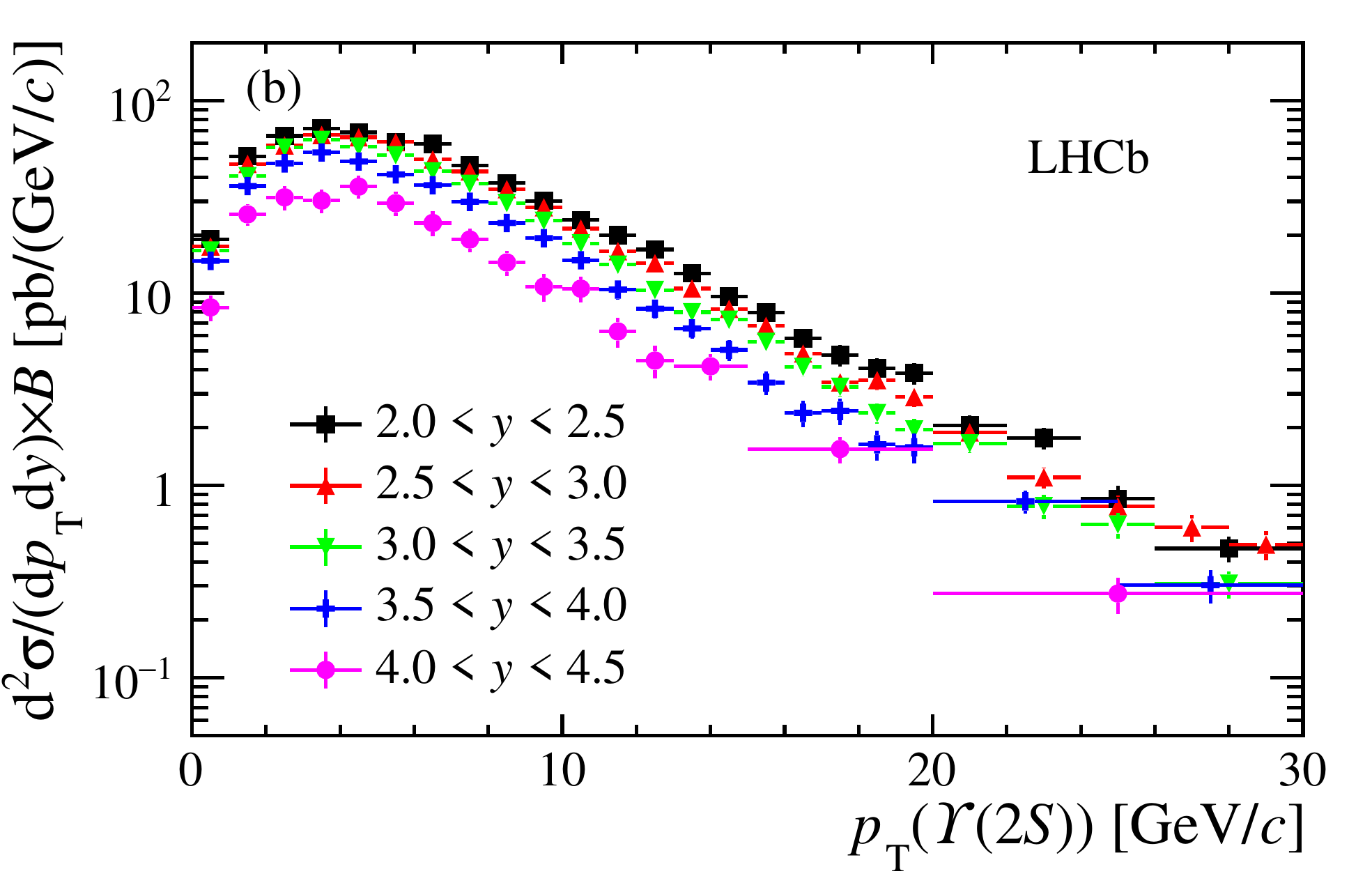} 
\end{minipage} 
\centering
\begin{minipage}[t]{0.6\textwidth}
\centering 
\includegraphics[width=1.0\textwidth]{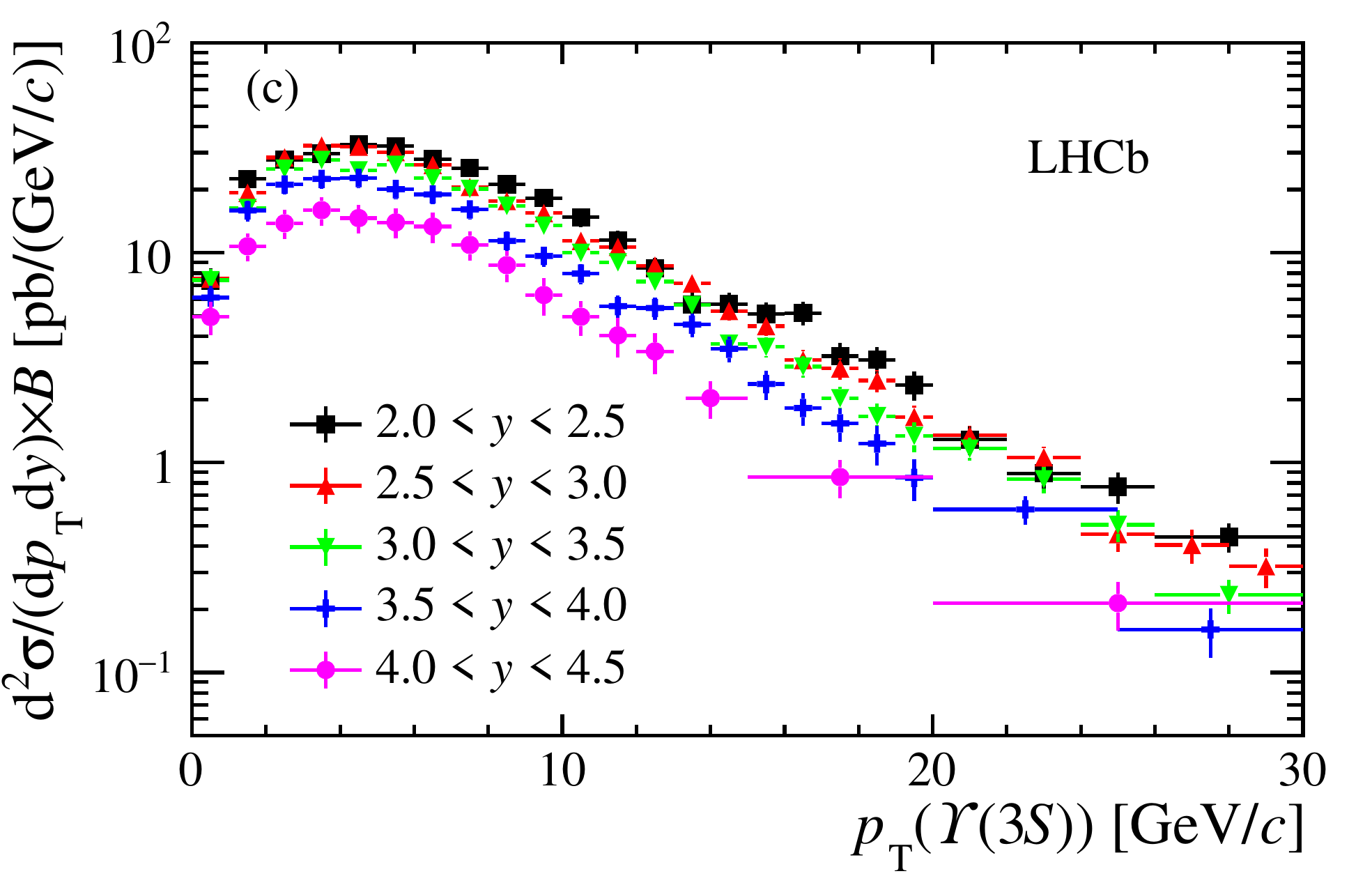}
\end{minipage} 
\caption{Double-differential cross-sections multiplied by dimuon branching fractions as a function of \pt in intervals of $y$ for the (a) $\OneS$, (b) $\TwoS$ and (c) $\ThreeS$ mesons. Statistical and systematic uncertainties are added in quadrature.}
\label{fig:Result1S}
\end{figure}

\begin{figure}[!tbp]
\centering
\begin{minipage}[t]{0.6\textwidth}
\centering
\includegraphics[width=1.0\textwidth]{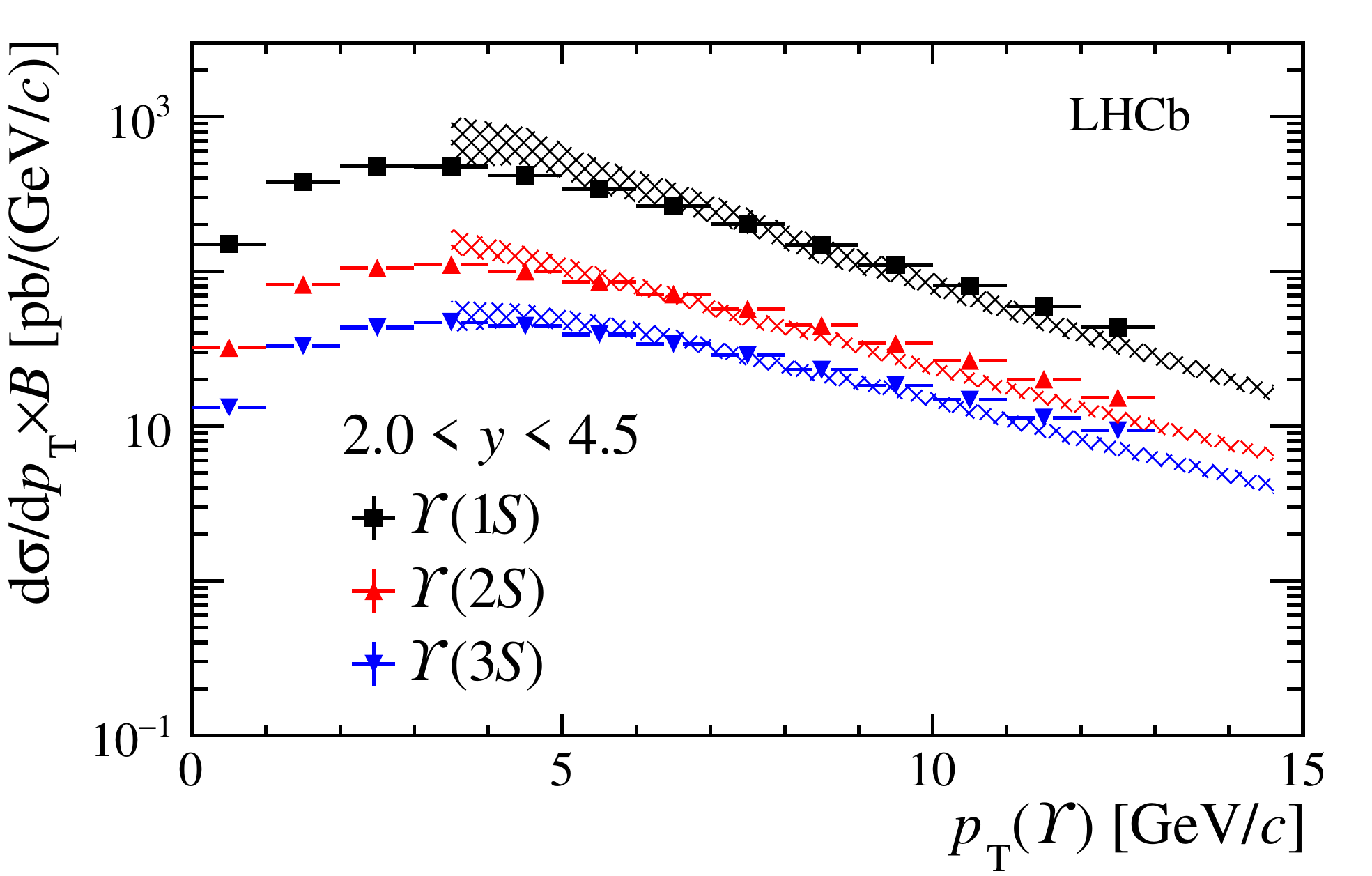}
\end{minipage}
\begin{minipage}[t]{0.6\textwidth}
\centering
\includegraphics[width=1.0\textwidth]{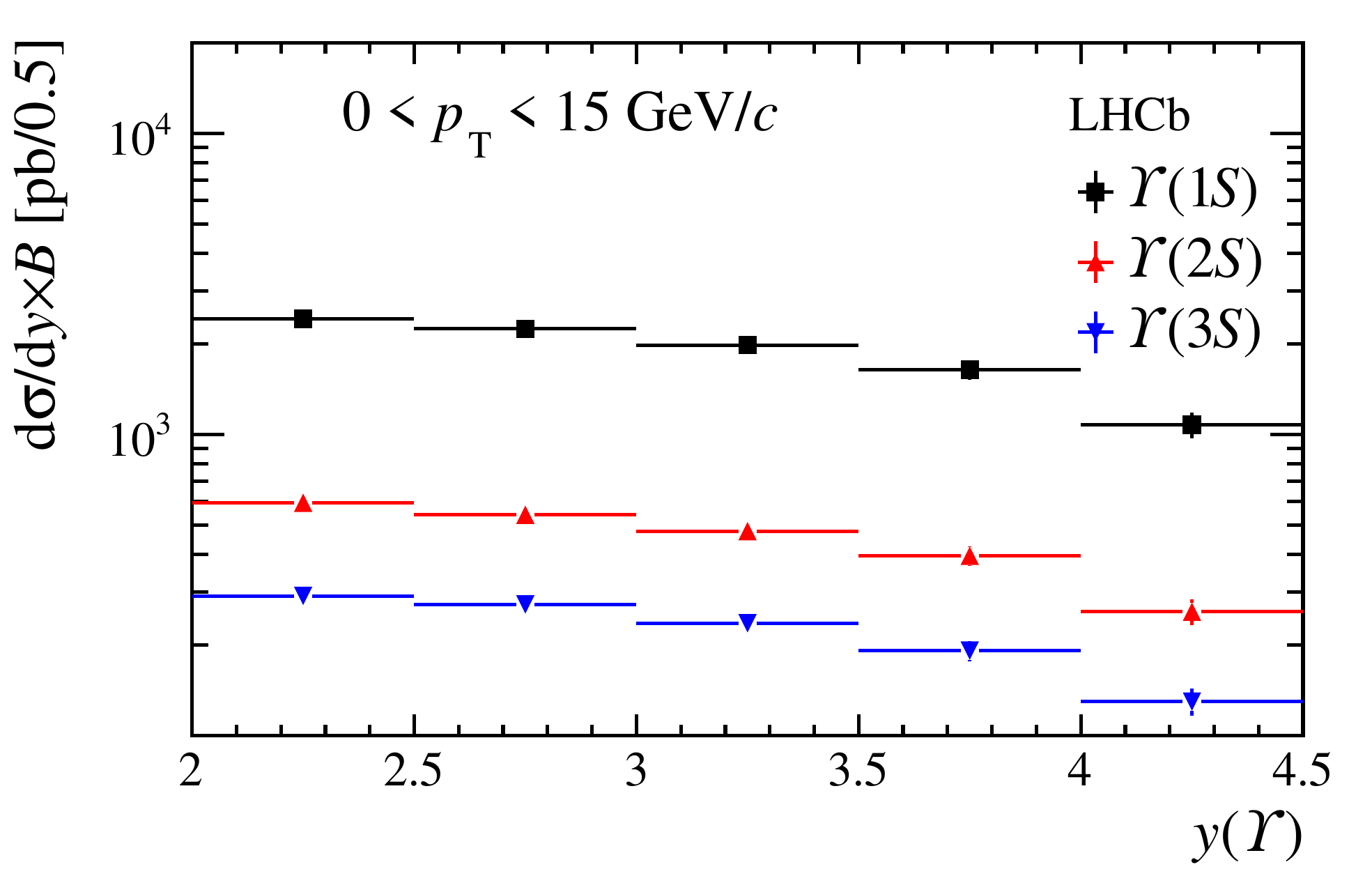}
\end{minipage}
\caption{Differential cross-sections multiplied by dimuon branching fractions 
for the $\OneS$ (black solid squares), $\TwoS$ (red upward triangles) and $\ThreeS$ (blue downward triangles) states (top) versus \pt 
integrated over $y$ between 2.0 and 4.5 and (bottom)
versus $y$ integrated over \pt from 0 to $15\gevc$. 
Statistical and systematic uncertainties are added in quadrature. 
Predictions from NRQCD~\cite{Wang:Upsilon2015} for the $\OneS$ (black grid
shading), $\TwoS$ (red grid shading) and $\ThreeS$ (blue grid shading) states
are overlaid in the top plot.}
\label{fig:Compare1S}
\end{figure}

The double-differential cross-sections multiplied by 
dimuon branching fractions for the $\OneS$, $\TwoS$ and $\ThreeS$ states are shown in 
Fig.~\ref{fig:Result1S}.  
The corresponding values are listed in 
Tables~\ref{tab:CS11}$-$\ref{tab:CS33}.
By integrating the double-differential results over $\pt\;(y)$,
the differential cross-sections times  
dimuon branching fractions as functions of $y\;(\pt)$ 
are shown in 
Fig.~\ref{fig:Compare1S} 
for the three $\ups$ states,
with the theoretical predictions based on
NRQCD~\cite{Wang:Upsilon2015} overlaid. 
The NRQCD predictions are in agreement with the experimental data at high $\pt$. 

\begin{figure}[!tbp]
\centering
\begin{minipage}[t]{0.6\textwidth}
\centering
\includegraphics[width=1.0\textwidth]{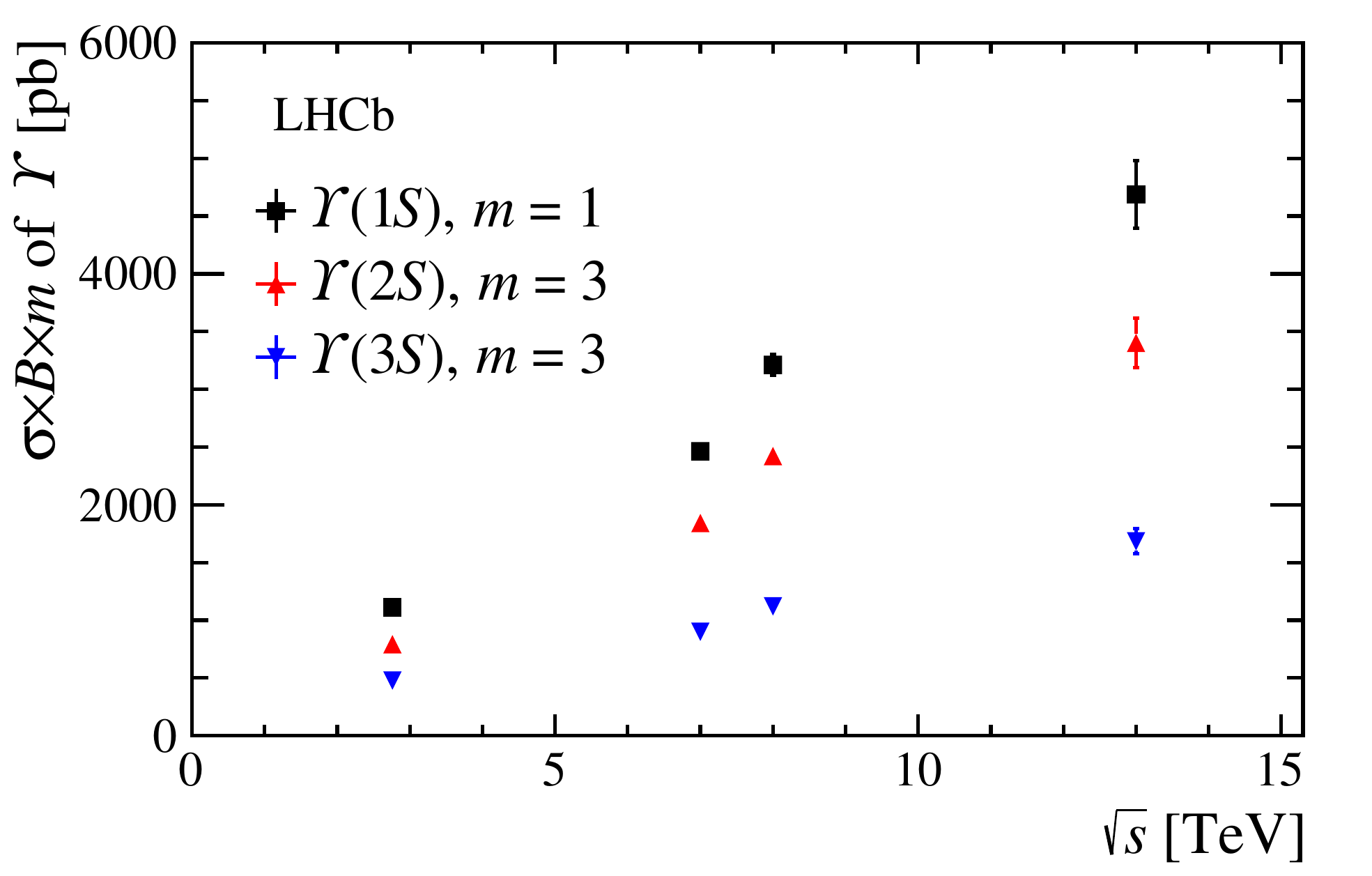}
\end{minipage}
\caption{The production cross-sections multiplied by dimuon branching fractions 
integrated over $0<\pt<15\gevc$ and $2.0<y<4.5$ versus centre-of-mass energy of $pp$ collisions for the $\OneS$ (black solid squares), $\TwoS$ (red upward triangles) and $\ThreeS$ (blue downward triangles) states. 
Each set of measurements is offset by a multiplicative factor $m$, which is shown on the plot. 
Statistical and systematic uncertainties are added in quadrature.}
\label{fig:CS123}
\end{figure}

The total cross-sections multiplied by dimuon branching fractions for
the three states integrated over the ranges of $0< \pt < 15\gevc$ and
$2.0< y < 4.5$ are measured to be
\begin{alignat*}{3}
\BR(\OneS\to\mumu)\times\sigma(\OneS, 0<\pt<15\gevc,2<y<4.5)     &= 4687 \pm 10  \pm 294~\pb,\nonumber\\
\BR(\TwoS\to\mumu)\times\sigma(\TwoS, 0<\pt<15\gevc,2<y<4.5)     &= 1134 \pm \phantom{0}6 \pm \phantom{0}71~\pb,\nonumber\\
\BR(\ThreeS\to\mumu)\times\sigma(\ThreeS, 0<\pt<15\gevc,2<y<4.5) &= \phantom{0}561 \pm \phantom{0}4 \pm \phantom{0}36~\pb,\nonumber
\end{alignat*}
where the first uncertainty is statistical and the second is systematic.
The corresponding results as a function of $pp$ centre-of-mass energy are shown 
in Fig.~\ref{fig:CS123}.

In this paper, results are
obtained under the assumption of zero polarisation.
The effects of possible $\ups$ polarisation 
based on the LHCb measurements~\cite{LHCb-PAPER-2017-028} in $pp$ collisions at 7$\tev$ and 8$\tev$ are studied.
The measurements show no large transverse or longitudinal polarisation over the accessible phase-space domain. 
The cross-sections increase up to $2.8\%$ for the three $\ups$ states when assuming the 
transverse polarisation of $\alpha=0.1$, where $\alpha$ is the polarisation parameter in the helicity frame~\cite{Wick:1959}.

\subsection{Cross-section ratios}
\begin{figure}[!tbp]
\centering
\begin{minipage}[t]{0.49\textwidth}
\centering
\includegraphics[width=1.0\textwidth]{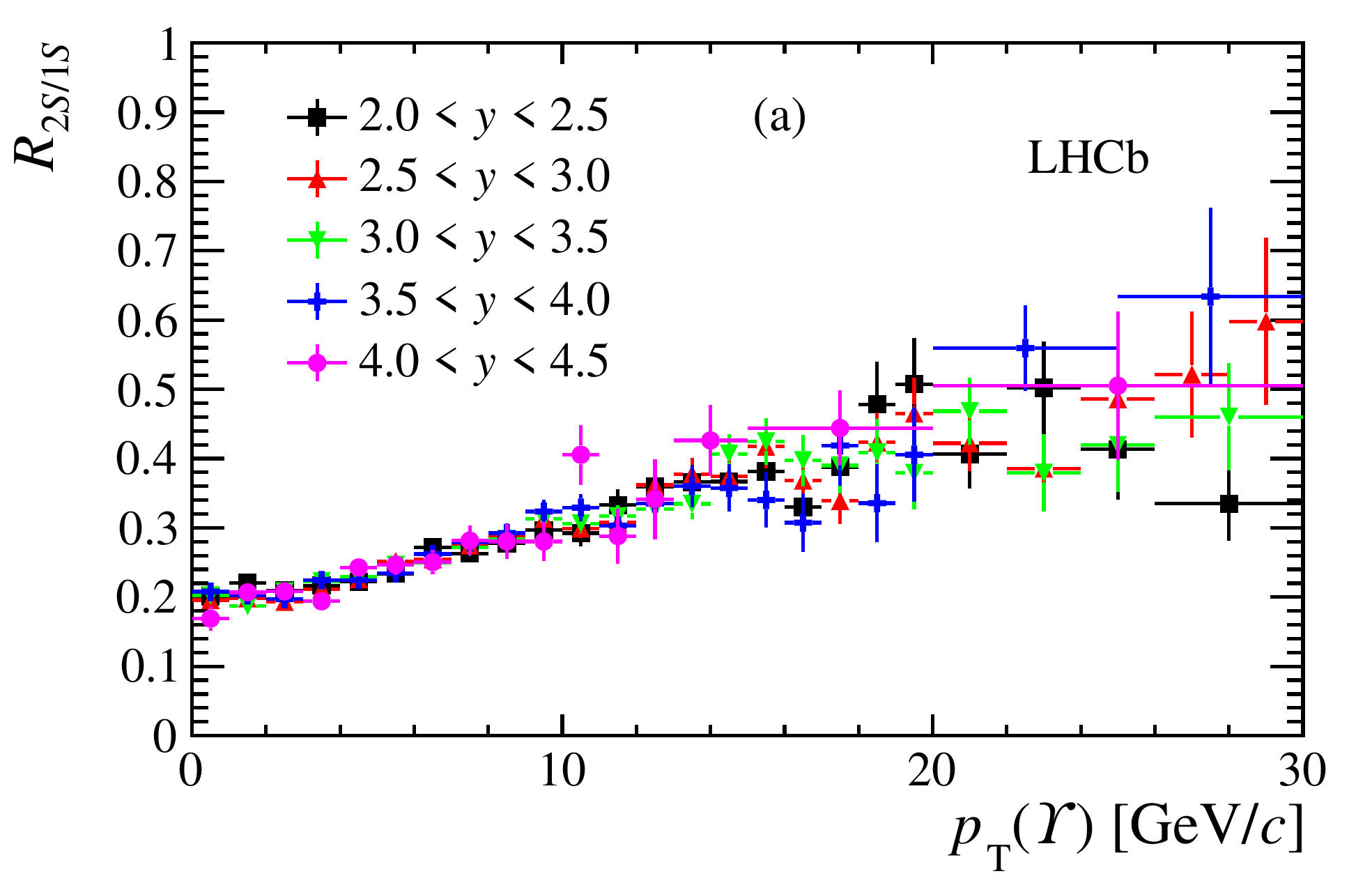}
\end{minipage}
\begin{minipage}[t]{0.49\textwidth}
\centering
\includegraphics[width=1.0\textwidth]{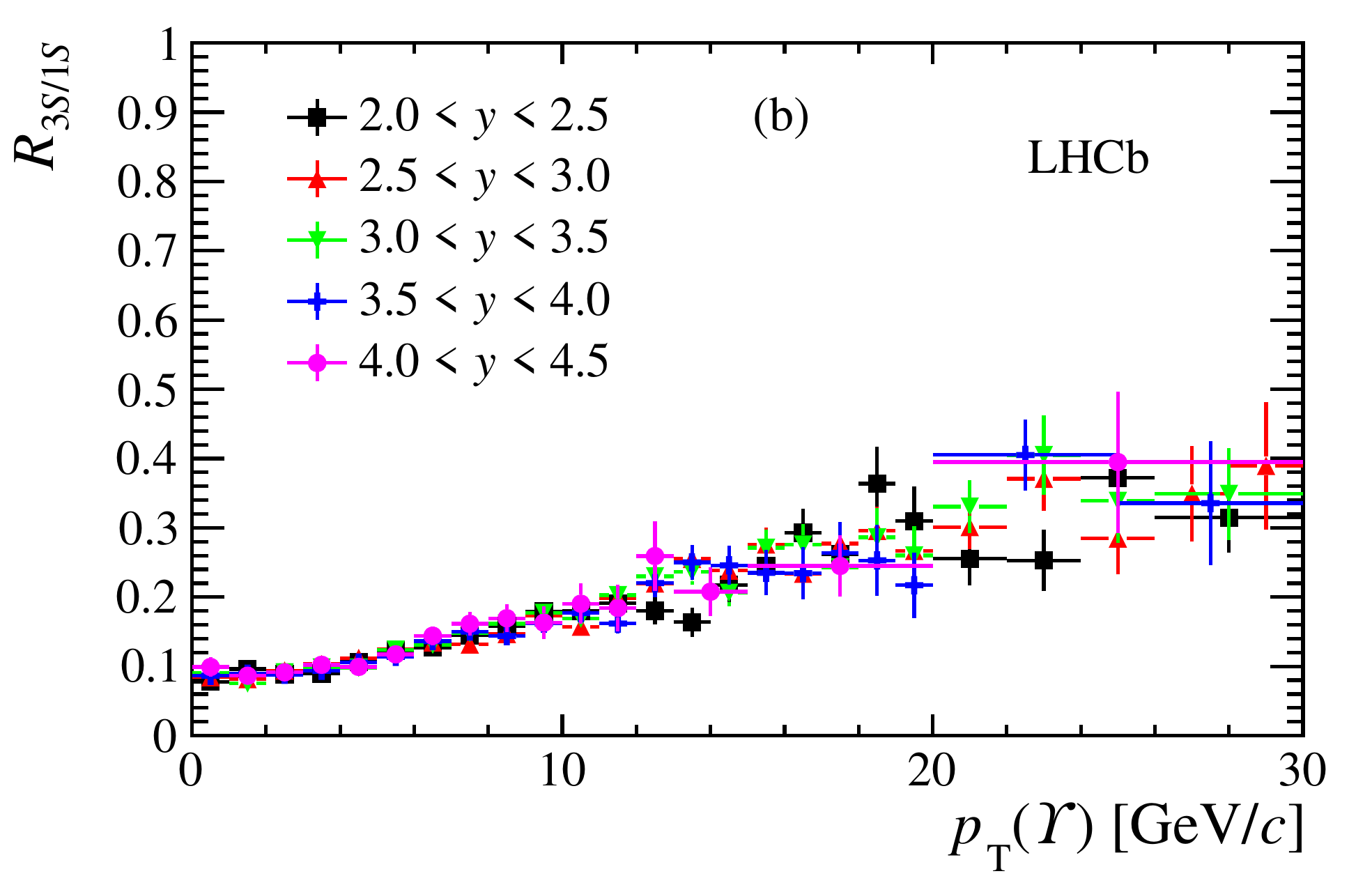}
\end{minipage}
\begin{minipage}[t]{0.49\textwidth}
\centering
\includegraphics[width=1.0\textwidth]{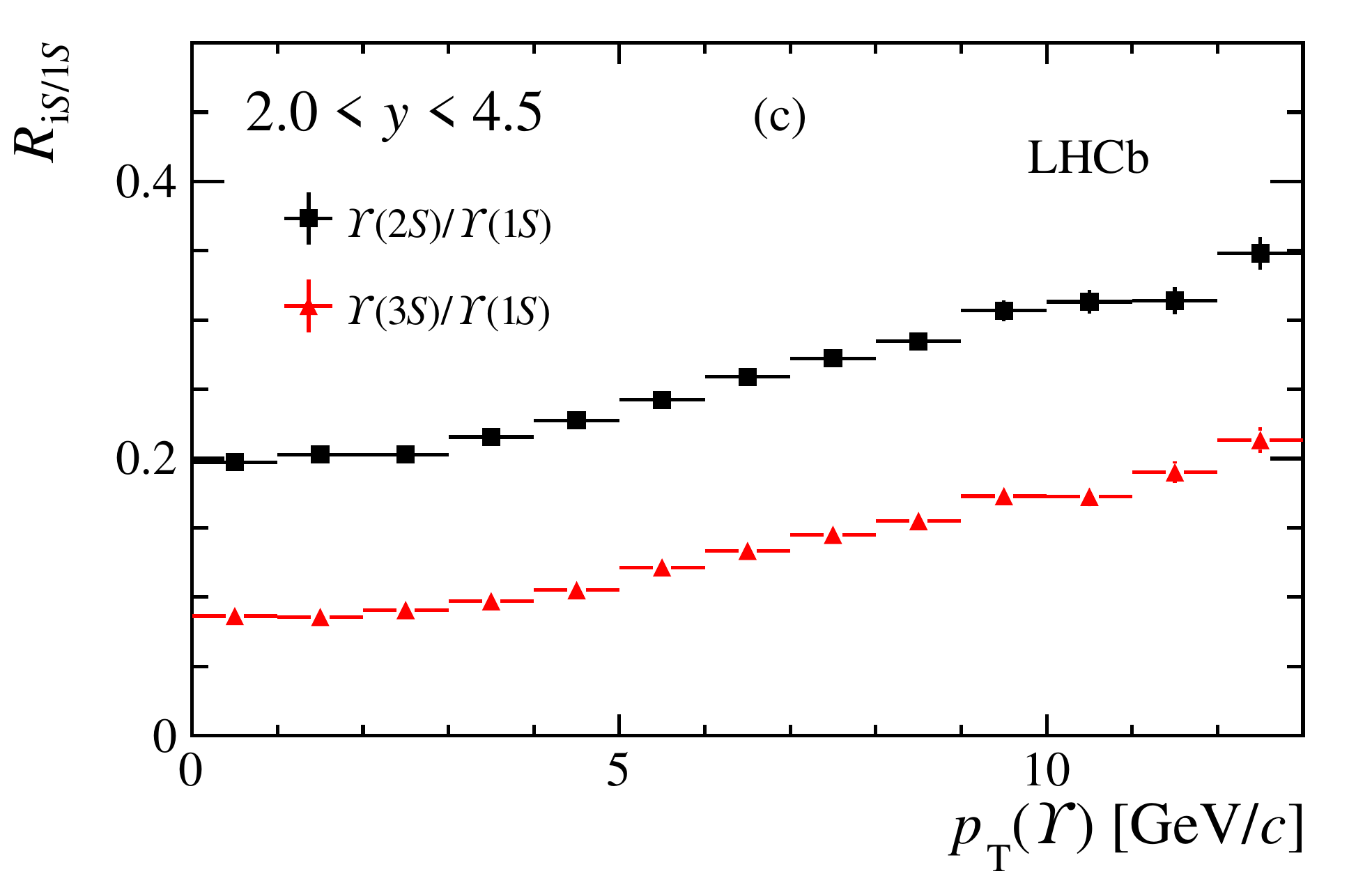}
\end{minipage}
\begin{minipage}[t]{0.49\textwidth}
\centering
\includegraphics[width=1.0\textwidth]{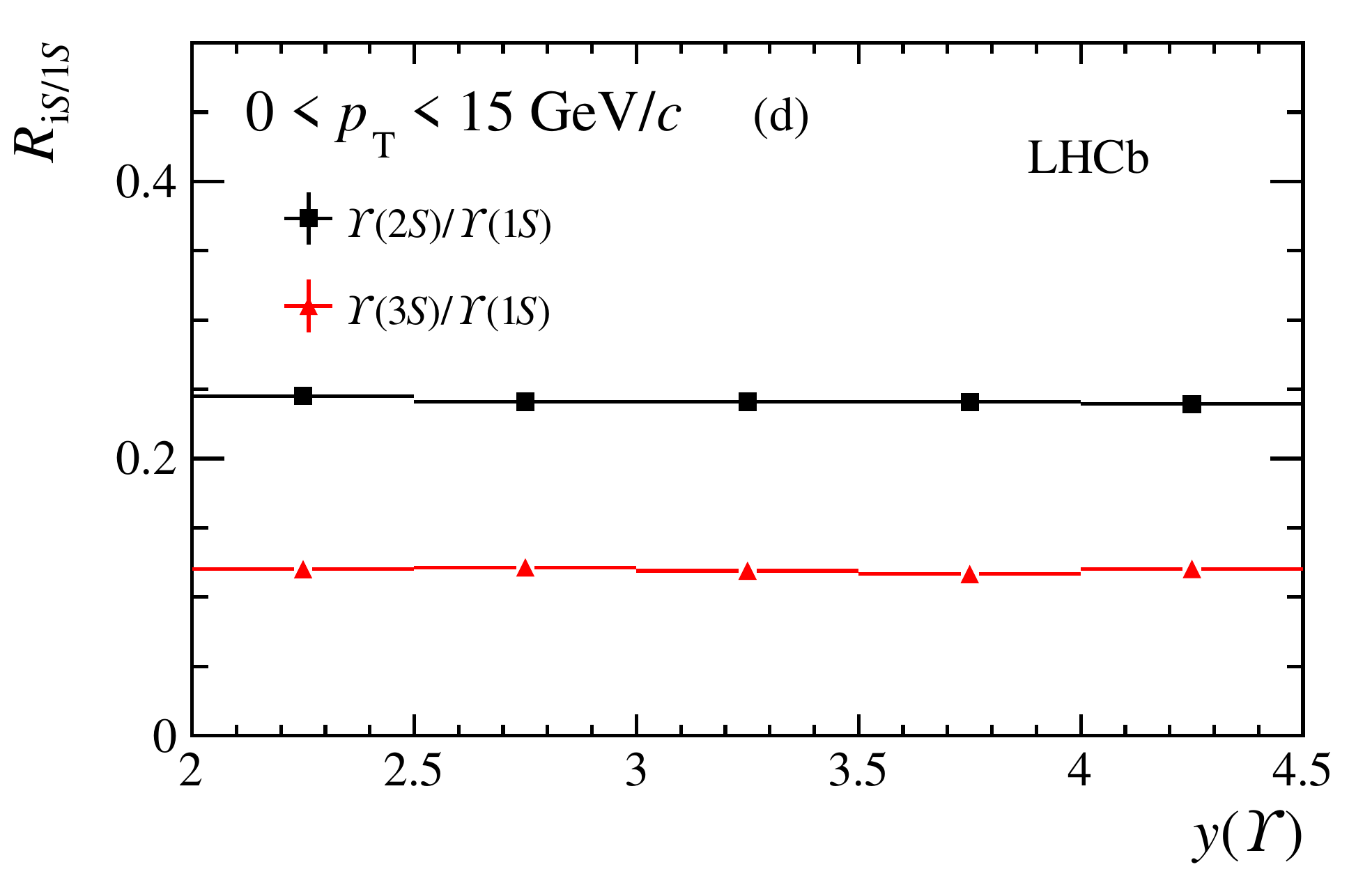}
\end{minipage}
\caption{
Ratios of double-differential cross-sections times dimuon branching fractions for (a) $\TwoS$ to $\OneS$ and (b) $\ThreeS$ to $\OneS$.
Ratios of differential cross-sections times dimuon branching fractions 
(c) versus $\pt$ integrated over $y$ and (d) versus $y$ integrated
over $\pt$ for $\TwoS$ to $\OneS$ (black solid squares) and $\ThreeS$
to $\OneS$ (red upward triangles).
Statistical and systematic uncertainties are added in quadrature.}
\label{fig:Ratio2131}
\end{figure}

The ratios of the differential production cross-sections
multiplied by dimuon branching fractions 
between $\TwoS$ and $\OneS$, $R_{2S/1S}$, and that between $\ThreeS$
and $\OneS$, $R_{3S/1S}$, are shown in 
Fig.~\ref{fig:Ratio2131}.

In these ratios, the statistical uncertainties of the cross-sections
and those due to the finite size of the simulated samples are assumed
to be uncorrelated.
The systematic uncertainties related to the signal yields and the efficiencies 
of the global event requirements, the trigger and the tracking are assumed to be 100$\%$ correlated between the different states. 

\begin{figure}[!tbp]
\centering
\begin{minipage}[t]{0.6\textwidth}
\centering
\includegraphics[width=1.0\textwidth]{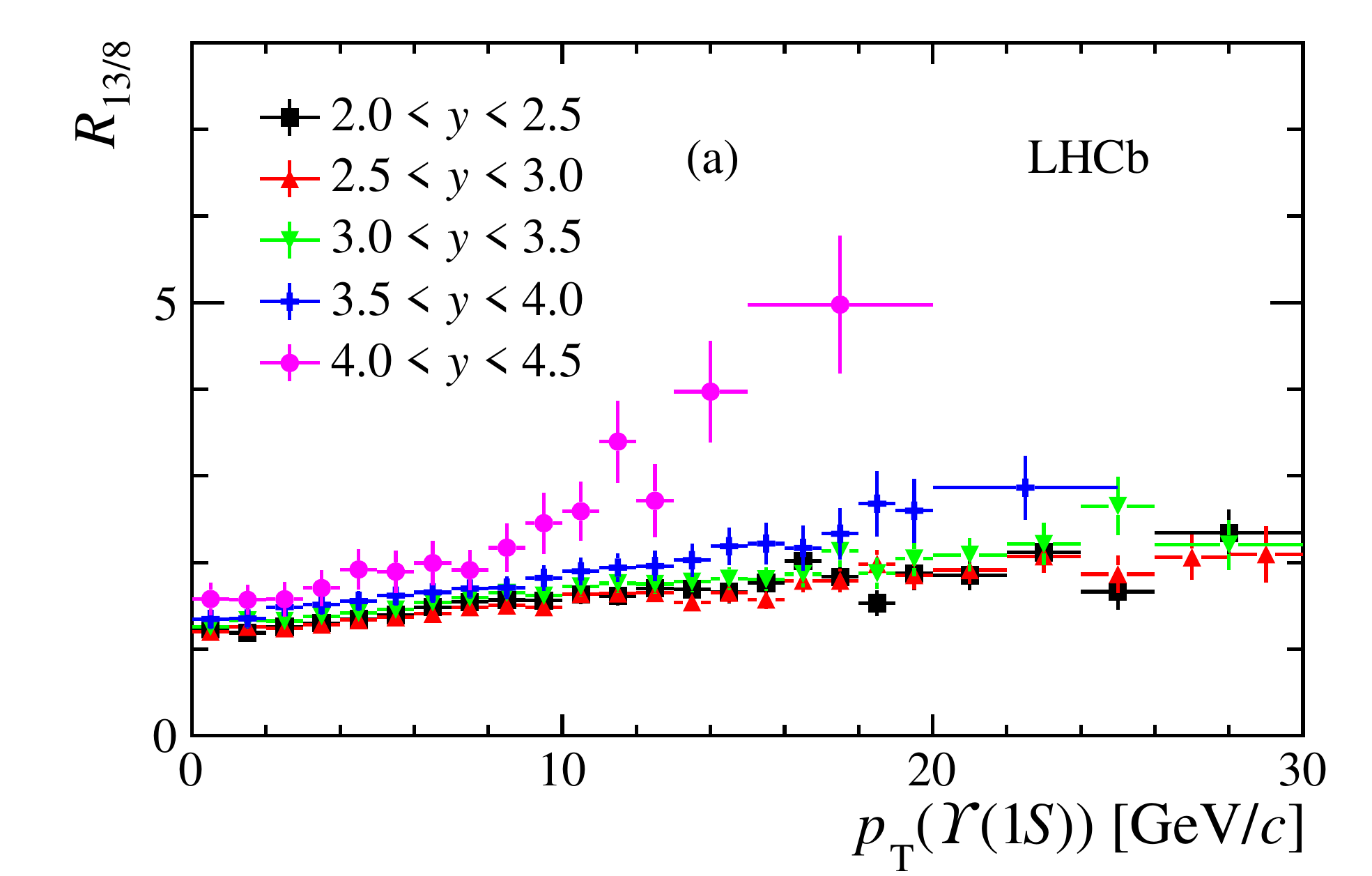}
\end{minipage}
\centering
\begin{minipage}[t]{0.6\textwidth}
\centering
\includegraphics[width=1.0\textwidth]{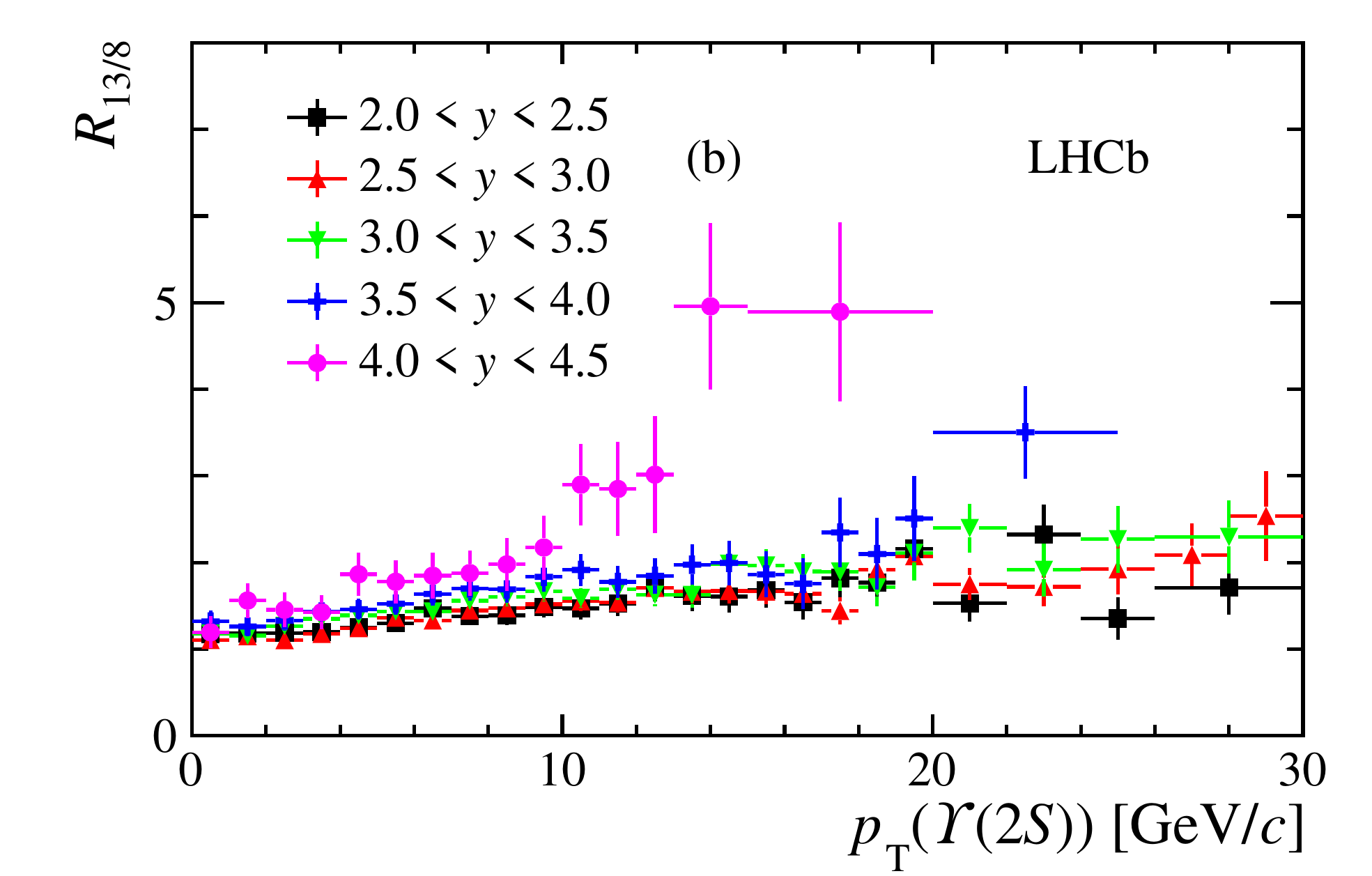}
\end{minipage}
\centering
\begin{minipage}[t]{0.6\textwidth}
\centering
\includegraphics[width=1.0\textwidth]{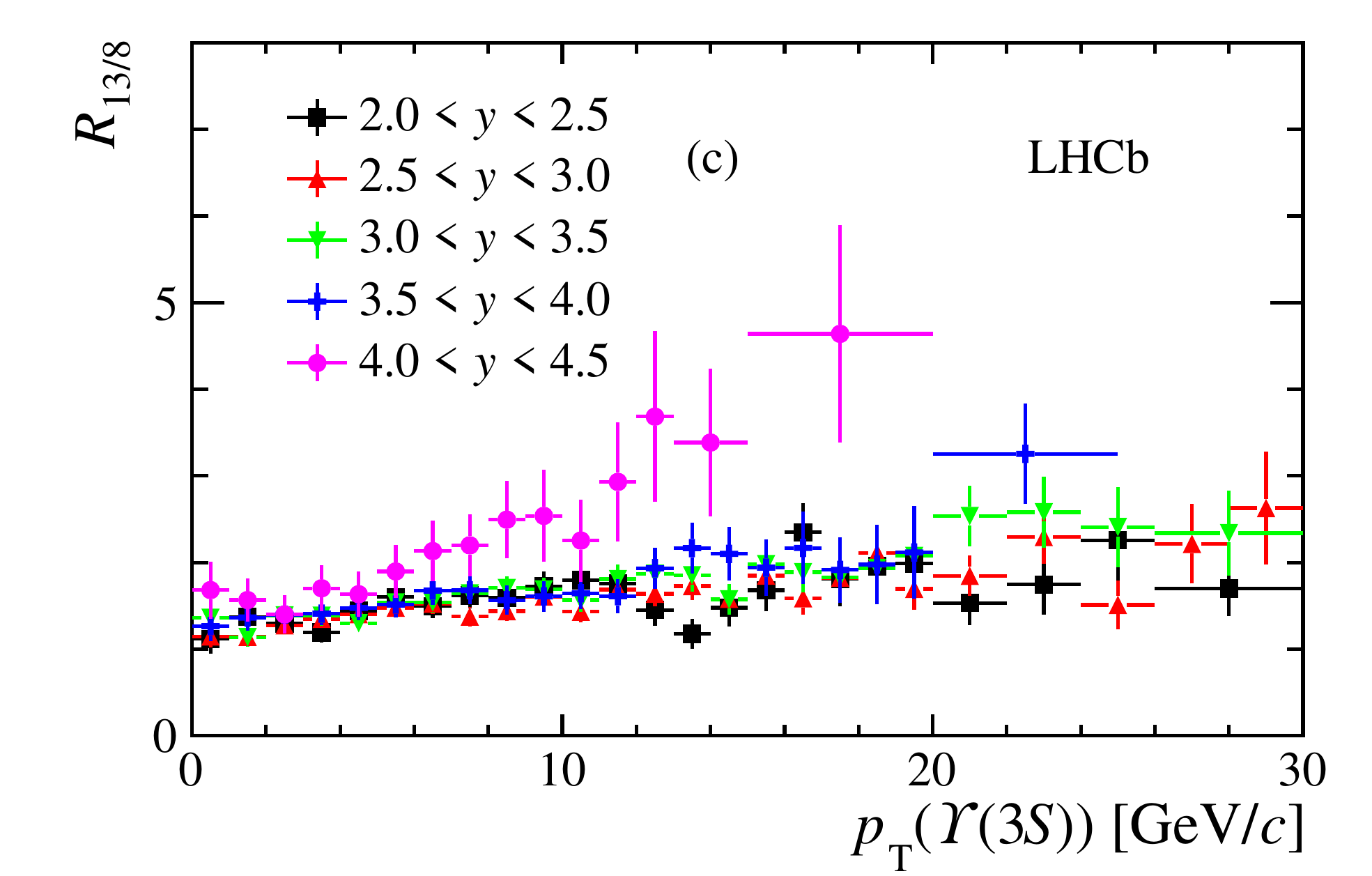}
\end{minipage}
\caption{Ratios of double-differential cross-sections between 13 TeV and 8 TeV measurements 
versus $\pt$ in intervals of $y$ for the (a) $\OneS$, (b) $\TwoS$ and (c) $\ThreeS$ states. 
Statistical and systematic uncertainties are added in quadrature.
}
\label{fig:RatioCS1}
\end{figure}

\begin{figure}[!tbp]
\centering
\begin{minipage}[t]{0.6\textwidth}
\centering
\includegraphics[width=1.0\textwidth]{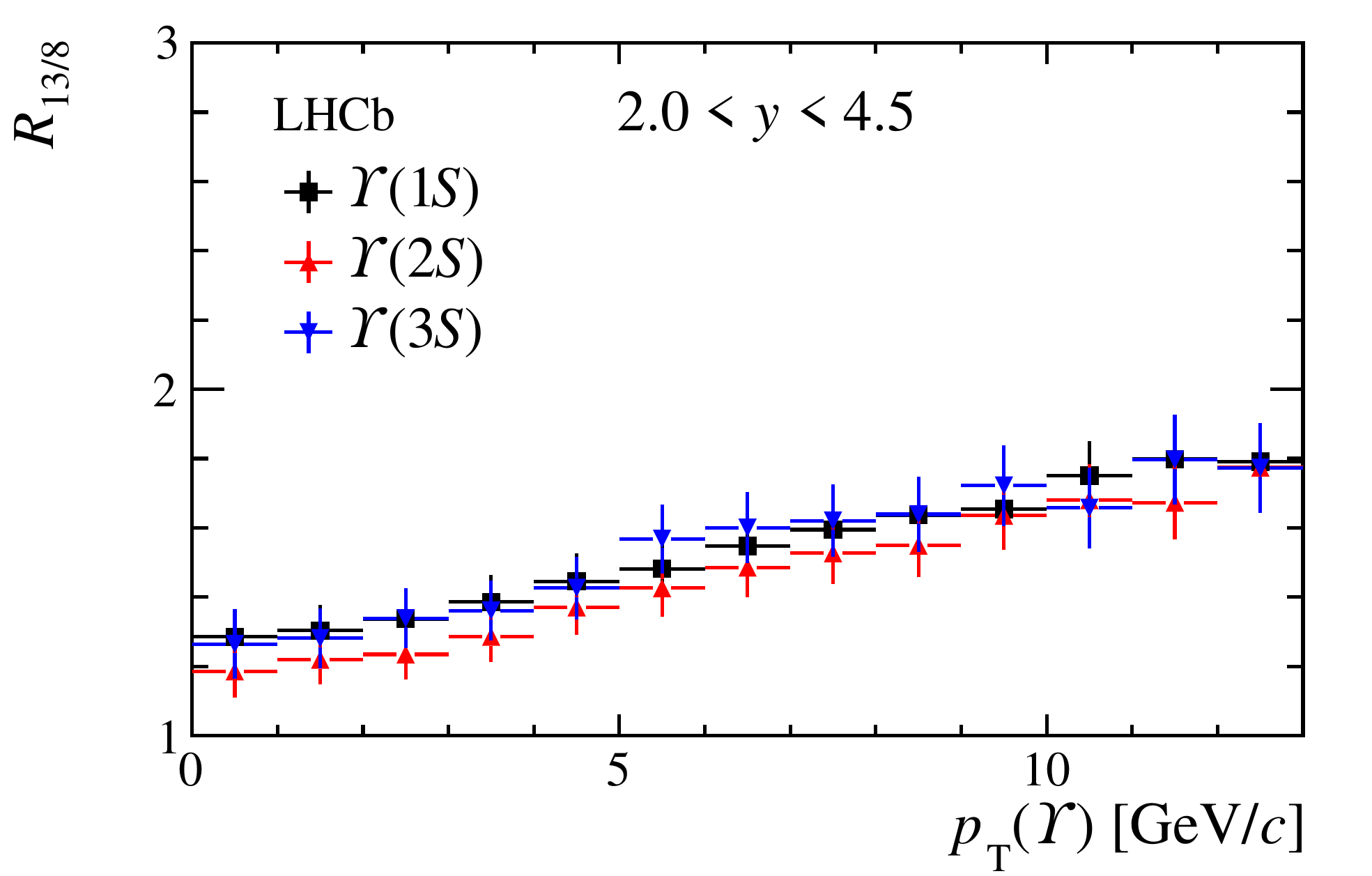}
\end{minipage}
\begin{minipage}[t]{0.6\textwidth}
\centering
\includegraphics[width=1.0\textwidth]{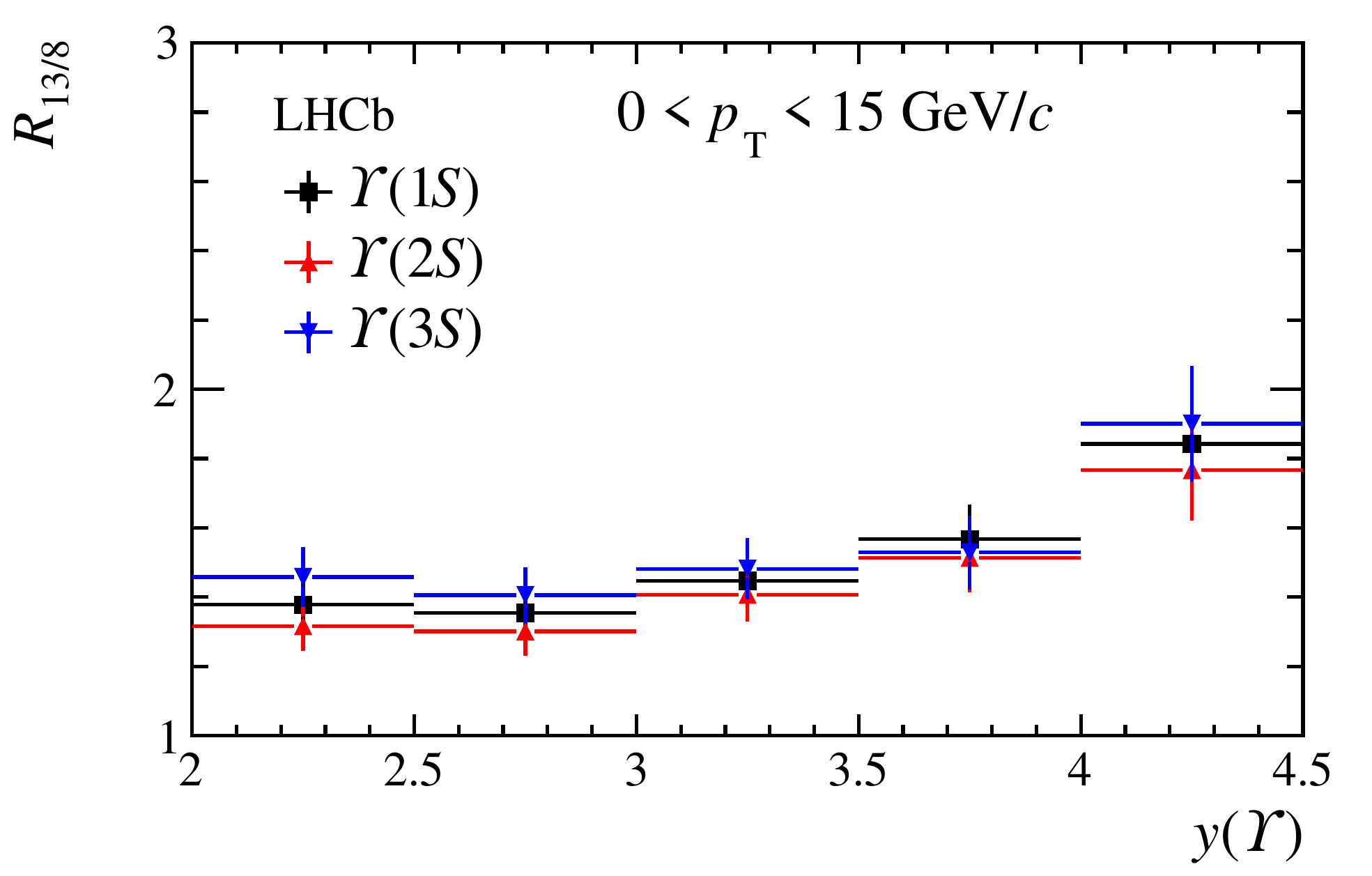}
\end{minipage}
\caption{Ratios of differential cross-sections between 13 TeV and 8 TeV measurements (top) versus $\pt$ integrated over $y$ and (bottom) versus $y$ integrated over $\pt$ for the $\OneS$ (black solid squares), $\TwoS$ (red upward triangles) and $\ThreeS$ (blue downward triangles) states. 
Statistical and systematic uncertainties are added in quadrature.}
\label{fig:RatioCSas1}
\end{figure}

The cross-sections times dimuon branching fractions 
measured at a centre-of-mass energy of $13\tev$ presented in this paper are compared with the measurements at $8\tev$~\cite{LHCb-PAPER-2015-045}.
The ratios of double-differential cross-sections between 13 TeV and 8 TeV measurements, $R_{13/8}$,
are shown in Fig.~\ref{fig:RatioCS1}.
The corresponding values are listed in Tables~\ref{tab:Ratio1381}$-$\ref{tab:Ratio1383}.
The cross-section ratios between $13\tev$ and $8\tev$ versus $\pt$ integrated over $y$, 
and versus $y$ integrated over $\pt$ are shown in 
Fig.~\ref{fig:RatioCSas1}.

In the ratios, the systematic uncertainties originating from the fit
model, global event requirements and kinematic spectrum are assumed to
be uncorrelated. 
Those from trigger, muon identification, tracking correction and luminosity are partially correlated.
The systematic uncertainty from final-state radiation is assumed to be $100\%$ correlated.

\section{Conclusion}\label{sec:conclusion}
A study of the production of $\OneS$, $\TwoS$ and $\ThreeS$ mesons in proton-proton collisions at the centre-of-mass energy $\sqs=13\tev$ is reported. 
The differential cross-sections times dimuon branching fractions of $\ups$ mesons are measured as functions of $\pt$ and $y$, in the kinematic range $0<\pt<30\gevc$ and $2.0<y<4.5$. 
The production cross-section ratios of $\TwoS$ and $\ThreeS$ mesons with respect to the $\OneS$ meson are given in intervals of $\pt$ and $y$. The ratios of the production cross-sections with respect to those measured at $\sqs=8\tev$ are also presented.

The results of differential cross-sections times dimuon branching fractions as a function of $\pt$ integrated over $y$ between 2.0 and 4.5 
are compared with predictions based on NRQCD. These predictions provide
a good description of the experimental data at high $\pt$ for all the three $\ups$ states.
\input{CS11}
\input{CS22}

\input{CS33}

\input{Ratio1381}

\input{Ratio1382}

\clearpage
\input{Ratio1383}

\clearpage

%% file: CS11.tex
\begin{sidewaystable}[!ht]
 \centering
 \begin{small}
   \begin{tabular*}{0.99\textwidth}{|@{\hspace{1mm}}c@{\extracolsep{\fill}}ccccc@{\hspace{1mm}}|}
      \hline
      $\pt\left[\!\gevc\right]$  
      &  $2.0<y<2.5$
      &  $2.5<y<3.0$
      &  $3.0<y<3.5$
      &  $3.5<y<4.0$
      &  $4.0<y<4.5$  
      \\
      \hline 
0$-$1
& $94.75 \pm 2.44  \pm 6.27$& $89.26 \pm 1.65  \pm 5.65$& $82.21 \pm 1.65  \pm 5.63$& $70.76 \pm 1.55  \pm 6.10$& $49.85 \pm 1.74  \pm 5.94$\\ 
1$-$2
& $233.67 \pm 3.83  \pm 15.52$& $236.95 \pm 2.72  \pm 14.94$& $216.85 \pm 2.61  \pm 14.87$& $177.87 \pm 2.47  \pm 16.12$& $124.19 \pm 2.68  \pm 14.27$\\ 
2$-$3
& $313.67 \pm 4.40  \pm 20.72$& $303.07 \pm 3.14  \pm 18.95$& $274.80 \pm 2.89  \pm 18.39$& $240.07 \pm 2.88  \pm 20.28$& $151.23 \pm 2.94  \pm 20.38$\\ 
3$-$4
& $330.78 \pm 4.51  \pm 22.11$& $313.85 \pm 3.22  \pm 19.58$& $281.00 \pm 2.90  \pm 18.17$& $240.17 \pm 2.84  \pm 18.96$& $156.12 \pm 3.04  \pm 20.26$\\ 
4$-$5
& $308.47 \pm 4.36  \pm 20.08$& $285.77 \pm 3.09  \pm 18.01$& $251.96 \pm 2.70  \pm 16.26$& $215.22 \pm 2.71  \pm 17.63$& $147.40 \pm 3.00  \pm 18.69$\\ 
5$-$6
& $261.01 \pm 4.05  \pm 17.44$& $242.35 \pm 2.86  \pm 15.04$& $211.58 \pm 2.50  \pm 13.80$& $176.79 \pm 2.43  \pm 13.91$& $118.96 \pm 2.68  \pm 16.00$\\ 
6$-$7
& $219.98 \pm 3.74  \pm 13.90$& $194.56 \pm 2.51  \pm 12.11$& $172.49 \pm 2.20  \pm 11.10$& $138.96 \pm 2.16  \pm 11.22$& $92.48 \pm 2.41  \pm 12.40$\\ 
7$-$8
& $175.08 \pm 3.34  \pm 11.48$& $156.34 \pm 2.18  \pm 9.76$& $136.22 \pm 1.90  \pm 8.62$& $107.01 \pm 1.87  \pm 8.15$& $67.50 \pm 2.05  \pm 8.26$\\ 
8$-$9
& $134.33 \pm 2.86  \pm 8.69$& $119.99 \pm 1.86  \pm 7.49$& $103.59 \pm 1.58  \pm 6.65$& $79.26 \pm 1.57  \pm 6.62$& $51.49 \pm 1.85  \pm 6.58$\\ 
9$-$10
& $101.78 \pm 2.46  \pm 6.58$& $89.18 \pm 1.54  \pm 5.63$& $76.04 \pm 1.34  \pm 4.98$& $59.62 \pm 1.33  \pm 5.06$& $38.61 \pm 1.59  \pm 5.52$\\
10$-$11
& $81.97 \pm 2.14  \pm 5.57$& $72.34 \pm 1.34  \pm 4.54$& $59.17 \pm 1.16  \pm 3.92$& $44.99 \pm 1.13  \pm 3.88$& $26.02 \pm 1.30  \pm 3.16$\\ 
11$-$12
& $60.06 \pm 1.78  \pm 4.12$& $53.65 \pm 1.14  \pm 3.38$& $44.40 \pm 0.99  \pm 2.96$& $34.43 \pm 0.97  \pm 2.98$& $21.94 \pm 1.17  \pm 2.79$\\ 
12$-$13
& $46.88 \pm 1.54  \pm 3.25$& $39.55 \pm 0.98  \pm 2.50$& $31.71 \pm 0.83  \pm 2.12$& $24.76 \pm 0.82  \pm 2.24$& $13.07 \pm 0.93  \pm 1.69$
\\ \cline{6-6}
13$-$14
& $34.64 \pm 1.30  \pm 2.44$& $28.00 \pm 0.79  \pm 1.78$& $23.81 \pm 0.71  \pm 1.61$& $18.20 \pm 0.70  \pm 1.67$
& \multirow{2}{*}{$9.77 \pm 0.53  \pm 1.24$ } \\
14$-$15
& $26.20 \pm 1.10  \pm 1.85$& $22.15 \pm 0.68  \pm 1.42$& $17.87 \pm 0.61  \pm 1.23$& $14.18 \pm 0.60  \pm 1.30$
& \\  \cline{6-6}
15$-$16
& $20.81 \pm 0.97  \pm 1.49$& $16.20 \pm 0.58  \pm 1.05$& $13.14 \pm 0.51  \pm 0.91$& $10.07 \pm 0.51  \pm 1.01$
& \multirow{5}{*}{$3.48 \pm 0.20  \pm 0.44$ } \\
16$-$17
& $17.64 \pm 0.87  \pm 1.27$& $13.17 \pm 0.53  \pm 0.87$& $10.41 \pm 0.46  \pm 0.75$& $7.76 \pm 0.46  \pm 0.81$
 &  \\
17$-$18
& $12.29 \pm 0.65  \pm 0.89$& $10.13 \pm 0.45  \pm 0.68$& $8.35 \pm 0.41  \pm 0.61$& $5.83 \pm 0.39  \pm 0.62$
 &  \\
18$-$19
& $8.50 \pm 0.53  \pm 0.63$& $8.33 \pm 0.42  \pm 0.58$& $5.82 \pm 0.35  \pm 0.44$& $4.88 \pm 0.37  \pm 0.54$
 &  \\
19$-$20
& $7.57 \pm 0.50  \pm 0.58$& $6.19 \pm 0.34  \pm 0.43$& $5.15 \pm 0.32  \pm 0.39$& $3.90 \pm 0.31  \pm 0.43$
 &  \\  \cline{2-6}
20$-$21
& \multirow{2}{*}{$5.04 \pm 0.29  \pm 0.37$ }  & \multirow{2}{*}{$4.47 \pm 0.21  \pm 0.31$ }  & \multirow{2}{*}{$3.53 \pm 0.18  \pm 0.27$ }  & \multirow{5}{*}{$1.48 \pm 0.09  \pm 0.17$ } & \multirow{10}{*}{$0.54 \pm 0.06  \pm 0.07$ } \\
21$-$22
&   &   &   &  &  \\  \cline{2-4}
22$-$23
& \multirow{2}{*}{$3.52 \pm 0.24  \pm 0.27$ }  & \multirow{2}{*}{$2.86 \pm 0.17  \pm 0.21$ }  & \multirow{2}{*}{$2.06 \pm 0.14  \pm 0.17$ }  &  &  \\
23$-$24
&   &   &   &  &  \\  \cline{2-4}
24$-$25
& \multirow{2}{*}{ $2.06 \pm 0.17  \pm 0.16$ }  & \multirow{2}{*}{$1.60 \pm 0.12  \pm 0.12$ }  & \multirow{2}{*}{$1.49 \pm 0.12  \pm 0.13$ }  &  &  \\  \cline{5-5}
25$-$26
&   &   &  & \multirow{5}{*}{$0.48 \pm 0.05  \pm 0.06$ } &  \\  \cline{2-4}
26$-$27
& \multirow{4}{*}{$1.40 \pm 0.10  \pm 0.11$ }  & \multirow{2}{*}{$1.16 \pm 0.10  \pm 0.10$ }  & \multirow{4}{*}{$0.67 \pm 0.06  \pm 0.06$ } &  &  \\
27$-$28
&   &   &  &  &  \\  \cline{3-3}
28$-$29
&   & \multirow{2}{*}{$0.82 \pm 0.09  \pm 0.07$ }  &  &  &  \\
29$-$30
 &   &   &  &  & \\
\hline 
\end{tabular*}
\caption{Double-differential cross-sections times dimuon branching fraction 
   in different bins of \pt and $y$ for $\OneS$ (in~\pb). The first uncertainty is statistical and the second is systematic.}
\label{tab:CS11}
\end{small}
\end{sidewaystable}

%% file: CS22.tex
\begin{sidewaystable}[!ht]
 \centering
 \begin{small}
   \begin{tabular*}{0.99\textwidth}{|@{\hspace{1mm}}c@{\extracolsep{\fill}}ccccc@{\hspace{1mm}}|}
      \hline
      $\pt\left[\!\gevc\right]$  
      &  $2.0<y<2.5$
      &  $2.5<y<3.0$
      &  $3.0<y<3.5$
      &  $3.5<y<4.0$
      &  $4.0<y<4.5$  
      \\
      \hline 
0$-$1
& $19.09 \pm 1.26  \pm 1.29$& $17.48 \pm 0.87  \pm 1.11$& $16.64 \pm 0.92  \pm 1.14$& $14.71 \pm 0.83  \pm 1.26$& $8.43 \pm 0.79  \pm 1.02$\\ 
1$-$2
& $51.52 \pm 2.06  \pm 3.43$& $47.02 \pm 1.45  \pm 2.96$& $40.56 \pm 1.42  \pm 2.77$& $36.04 \pm 1.33  \pm 3.25$& $25.69 \pm 1.38  \pm 3.04$\\ 
2$-$3
& $65.75 \pm 2.33  \pm 4.33$& $58.61 \pm 1.65  \pm 3.66$& $57.16 \pm 1.63  \pm 3.85$& $47.30 \pm 1.52  \pm 4.04$& $31.50 \pm 1.49  \pm 4.31$\\ 
3$-$4
& $71.64 \pm 2.44  \pm 4.76$& $66.47 \pm 1.78  \pm 4.14$& $62.67 \pm 1.65  \pm 4.06$& $53.90 \pm 1.59  \pm 4.38$& $30.28 \pm 1.52  \pm 4.01$\\ 
4$-$5
& $68.54 \pm 2.36  \pm 4.43$& $64.62 \pm 1.75  \pm 4.06$& $57.78 \pm 1.53  \pm 3.72$& $48.37 \pm 1.50  \pm 4.04$& $35.79 \pm 1.65  \pm 4.62$\\ 
5$-$6
& $61.05 \pm 2.23  \pm 4.05$& $61.03 \pm 1.68  \pm 3.78$& $52.15 \pm 1.45  \pm 3.39$& $41.41 \pm 1.37  \pm 3.31$& $29.36 \pm 1.54  \pm 3.96$\\ 
6$-$7
& $59.77 \pm 2.22  \pm 3.78$& $49.55 \pm 1.50  \pm 3.08$& $43.04 \pm 1.29  \pm 2.76$& $36.43 \pm 1.26  \pm 2.98$& $23.17 \pm 1.39  \pm 3.17$\\ 
7$-$8
& $46.04 \pm 1.94  \pm 3.03$& $42.95 \pm 1.32  \pm 2.68$& $37.01 \pm 1.12  \pm 2.34$& $29.84 \pm 1.14  \pm 2.33$& $19.00 \pm 1.25  \pm 2.35$\\ 
8$-$9
& $37.33 \pm 1.76  \pm 2.43$& $34.72 \pm 1.14  \pm 2.16$& $29.41 \pm 0.95  \pm 1.88$& $23.17 \pm 0.97  \pm 1.99$& $14.44 \pm 1.10  \pm 1.91$\\ 
9$-$10
& $30.18 \pm 1.52  \pm 1.98$& $27.93 \pm 0.98  \pm 1.76$& $23.81 \pm 0.81  \pm 1.57$& $19.29 \pm 0.84  \pm 1.64$& $10.82 \pm 0.92  \pm 1.57$\\ 
10$-$11
& $23.97 \pm 1.32  \pm 1.66$& $21.65 \pm 0.83  \pm 1.37$& $18.12 \pm 0.71  \pm 1.20$& $14.81 \pm 0.71  \pm 1.27$& $10.55 \pm 0.91  \pm 1.33$\\ 
11$-$12
& $19.99 \pm 1.17  \pm 1.41$& $16.51 \pm 0.69  \pm 1.04$& $14.06 \pm 0.60  \pm 0.94$& $10.44 \pm 0.59  \pm 0.93$& $6.32 \pm 0.76  \pm 0.81$\\ 
12$-$13
& $16.85 \pm 1.03  \pm 1.19$& $14.30 \pm 0.61  \pm 0.90$& $10.38 \pm 0.51  \pm 0.71$& $8.30 \pm 0.52  \pm 0.75$& $4.46 \pm 0.64  \pm 0.57$
\\ \cline{6-6}
13$-$14
& $12.67 \pm 0.94  \pm 0.92$& $10.57 \pm 0.53  \pm 0.67$& $7.96 \pm 0.44  \pm 0.54$& $6.55 \pm 0.44  \pm 0.62$
& \multirow{2}{*}{ $4.16 \pm 0.40  \pm 0.53$ } \\
14$-$15
& $9.61 \pm 0.75  \pm 0.71$& $8.29 \pm 0.46  \pm 0.53$& $7.27 \pm 0.41  \pm 0.51$& $5.07 \pm 0.39  \pm 0.47$
& \\  \cline{6-6}
15$-$16
& $7.93 \pm 0.66  \pm 0.58$& $6.76 \pm 0.40  \pm 0.44$& $5.58 \pm 0.35  \pm 0.39$& $3.43 \pm 0.34  \pm 0.35$
& \multirow{5}{*}{$1.55 \pm 0.15  \pm 0.19$ } \\
16$-$17
& $5.82 \pm 0.55  \pm 0.43$& $4.85 \pm 0.34  \pm 0.32$& $4.14 \pm 0.31  \pm 0.30$& $2.39 \pm 0.27  \pm 0.25$
 &  \\
17$-$18
& $4.77 \pm 0.48  \pm 0.37$& $3.43 \pm 0.28  \pm 0.23$& $3.26 \pm 0.27  \pm 0.24$& $2.44 \pm 0.27  \pm 0.27$
 &  \\
18$-$19
& $4.07 \pm 0.42  \pm 0.31$& $3.53 \pm 0.28  \pm 0.24$& $2.38 \pm 0.23  \pm 0.18$& $1.64 \pm 0.23  \pm 0.18$
 &  \\
19$-$20
& $3.84 \pm 0.39  \pm 0.31$& $2.88 \pm 0.25  \pm 0.20$& $1.95 \pm 0.22  \pm 0.15$& $1.58 \pm 0.20  \pm 0.19$
 &  \\  \cline{2-6}
20$-$21
& \multirow{2}{*}{$2.05 \pm 0.20  \pm 0.16$ }  & \multirow{2}{*}{ $1.89 \pm 0.14  \pm 0.13$ }  & \multirow{2}{*}{$1.65 \pm 0.13  \pm 0.13$}& \multirow{5}{*}{$0.83 \pm 0.07  \pm 0.09$ } &\multirow{10}{*}{$0.27 \pm 0.05  \pm 0.04$ } \\
21$-$22
&   &   &   &  &  \\  \cline{2-4}
22$-$23
& \multirow{2}{*}{$1.77 \pm 0.18  \pm 0.14$ }  & \multirow{2}{*}{ $1.10 \pm 0.11  \pm 0.08$ }  & \multirow{2}{*}{$0.78 \pm 0.09  \pm 0.06$ }  &  &  \\
23$-$24
&   &   &   &  &  \\  \cline{2-4}
24$-$25
& \multirow{2}{*}{ $0.85 \pm 0.12  \pm 0.07$ }  & \multirow{2}{*}{$0.78 \pm 0.09  \pm 0.06$ }  & \multirow{2}{*}{$0.63 \pm 0.08  \pm 0.05$ }  &  &  \\  \cline{5-5}
25$-$26
&   &   &  & \multirow{5}{*}{$0.30 \pm 0.05  \pm 0.04$ } &  \\  \cline{2-4}
26$-$27
& \multirow{4}{*}{$0.47 \pm 0.06  \pm 0.04$ }  & \multirow{2}{*}{$0.61 \pm 0.08  \pm 0.05$ }  & \multirow{4}{*}{$0.31 \pm 0.04  \pm 0.03$ } &  &  \\
27$-$28
&   &   &  &  &  \\  \cline{3-3}
28$-$29
&   & \multirow{2}{*}{ $0.49 \pm 0.07  \pm 0.04$ }  &  &  &  \\
29$-$30
 &   &   &  &  & \\\hline 
\end{tabular*}
 \caption{Double-differential cross-sections times dimuon branching
   fraction in different bins of \pt and $y$ for $\TwoS$ (in~\pb). The first uncertainty is statistical and the second is systematic.}
   \label{tab:CS22}
 \end{small}
\end{sidewaystable}

%% file: CS33.tex
\begin{sidewaystable}[!ht]
 \centering
 \begin{small}
   \begin{tabular*}{0.99\textwidth}{|@{\hspace{1mm}}c@{\extracolsep{\fill}}ccccc@{\hspace{1mm}}|}
      \hline
      $\pt\left[\!\gevc\right]$  
      &  $2.0<y<2.5$
      &  $2.5<y<3.0$
      &  $3.0<y<3.5$
      &  $3.5<y<4.0$
      &  $4.0<y<4.5$  
      \\
      \hline 
0$-$1
& $7.37 \pm 0.91  \pm 0.51$& $7.54 \pm 0.66  \pm 0.50$& $7.41 \pm 0.78  \pm 0.52$& $6.12 \pm 0.64  \pm 0.55$& $4.94 \pm 0.67  \pm 0.62$\\ 
1$-$2
& $22.47 \pm 1.58  \pm 1.54$& $19.32 \pm 1.10  \pm 1.26$& $16.34 \pm 1.13  \pm 1.16$& $15.91 \pm 1.05  \pm 1.48$& $10.74 \pm 0.96  \pm 1.35$\\ 
2$-$3
& $27.76 \pm 1.75  \pm 1.90$& $28.45 \pm 1.32  \pm 1.84$& $25.14 \pm 1.32  \pm 1.74$& $21.13 \pm 1.21  \pm 1.85$& $13.83 \pm 1.12  \pm 1.92$\\ 
3$-$4
& $29.69 \pm 1.81  \pm 2.04$& $32.38 \pm 1.42  \pm 2.09$& $27.69 \pm 1.31  \pm 1.86$& $22.49 \pm 1.24  \pm 1.88$& $15.97 \pm 1.22  \pm 2.17$\\ 
4$-$5
& $32.80 \pm 1.85  \pm 2.20$& $32.05 \pm 1.42  \pm 2.09$& $24.71 \pm 1.19  \pm 1.65$& $22.72 \pm 1.19  \pm 1.94$& $14.63 \pm 1.21  \pm 1.91$\\ 
5$-$6
& $32.21 \pm 1.81  \pm 2.21$& $30.17 \pm 1.34  \pm 1.94$& $26.25 \pm 1.16  \pm 1.77$& $20.10 \pm 1.10  \pm 1.69$& $13.96 \pm 1.18  \pm 1.96$\\ 
6$-$7
& $27.96 \pm 1.74  \pm 1.82$& $26.21 \pm 1.23  \pm 1.69$& $22.66 \pm 1.06  \pm 1.51$& $18.97 \pm 1.03  \pm 1.61$& $13.34 \pm 1.16  \pm 1.87$\\ 
7$-$8
& $25.34 \pm 1.62  \pm 1.71$& $20.59 \pm 1.05  \pm 1.33$& $20.17 \pm 0.90  \pm 1.32$& $16.09 \pm 0.93  \pm 1.28$& $10.90 \pm 1.05  \pm 1.38$\\ 
8$-$9
& $21.21 \pm 1.45  \pm 1.42$& $17.63 \pm 0.92  \pm 1.14$& $16.74 \pm 0.78  \pm 1.11$& $11.39 \pm 0.77  \pm 1.01$& $8.73 \pm 0.96  \pm 1.14$\\ 
9$-$10
& $18.24 \pm 1.27  \pm 1.22$& $15.46 \pm 0.81  \pm 1.01$& $13.48 \pm 0.67  \pm 0.92$& $9.65 \pm 0.65  \pm 0.85$& $6.29 \pm 0.86  \pm 0.94$\\ 
10$-$11
& $14.77 \pm 1.17  \pm 1.04$& $11.39 \pm 0.67  \pm 0.74$& $9.99 \pm 0.57  \pm 0.69$& $7.96 \pm 0.57  \pm 0.70$& $4.96 \pm 0.69  \pm 0.62$\\ 
11$-$12
& $11.48 \pm 0.99  \pm 0.82$& $10.66 \pm 0.59  \pm 0.70$& $9.00 \pm 0.52  \pm 0.62$& $5.57 \pm 0.46  \pm 0.49$& $4.04 \pm 0.69  \pm 0.52$\\ 
12$-$13
& $8.45 \pm 0.84  \pm 0.61$& $8.69 \pm 0.53  \pm 0.57$& $7.29 \pm 0.44  \pm 0.51$& $5.45 \pm 0.43  \pm 0.49$& $3.39 \pm 0.59  \pm 0.46$
\\ \cline{6-6}
13$-$14
& $5.67 \pm 0.69  \pm 0.41$& $7.15 \pm 0.45  \pm 0.47$& $5.63 \pm 0.38  \pm 0.40$& $4.55 \pm 0.40  \pm 0.43$
& \multirow{2}{*}{$2.03 \pm 0.32  \pm 0.26$ } \\
14$-$15
& $5.70 \pm 0.64  \pm 0.42$& $5.28 \pm 0.38  \pm 0.35$& $3.68 \pm 0.31  \pm 0.26$& $3.49 \pm 0.35  \pm 0.34$
& \\  \cline{6-6}
15$-$16
& $5.11 \pm 0.57  \pm 0.38$& $4.47 \pm 0.35  \pm 0.30$& $3.56 \pm 0.30  \pm 0.26$& $2.37 \pm 0.29  \pm 0.24$
& \multirow{5}{*}{$0.85 \pm 0.14  \pm 0.11$ } \\
16$-$17
& $5.17 \pm 0.53  \pm 0.39$& $3.08 \pm 0.29  \pm 0.21$& $2.87 \pm 0.27  \pm 0.21$& $1.82 \pm 0.25  \pm 0.20$
 &  \\
17$-$18
& $3.22 \pm 0.44  \pm 0.25$& $2.81 \pm 0.26  \pm 0.19$& $2.03 \pm 0.23  \pm 0.15$& $1.54 \pm 0.23  \pm 0.16$
 &  \\
18$-$19
& $3.09 \pm 0.38  \pm 0.25$& $2.46 \pm 0.24  \pm 0.17$& $1.67 \pm 0.21  \pm 0.13$& $1.23 \pm 0.22  \pm 0.14$
 &  \\
19$-$20
& $2.34 \pm 0.33  \pm 0.19$& $1.65 \pm 0.19  \pm 0.12$& $1.34 \pm 0.19  \pm 0.11$& $0.85 \pm 0.16  \pm 0.10$
 &  \\  \cline{2-6}
20$-$21
& \multirow{2}{*}{ $1.29 \pm 0.17  \pm 0.10$ }  & \multirow{2}{*}{$1.35 \pm 0.12  \pm 0.09$ }  & \multirow{2}{*}{$1.17 \pm 0.11  \pm 0.09$ }  & \multirow{5}{*}{$0.60 \pm 0.06  \pm 0.07$ } & \multirow{10}{*}{$0.21 \pm 0.05  \pm 0.03$ } \\
21$-$22
&   &   &   &  &  \\  \cline{2-4}
22$-$23
& \multirow{2}{*}{$0.89 \pm 0.14  \pm 0.07$ }  & \multirow{2}{*}{$1.06 \pm 0.11  \pm 0.08$ }  & \multirow{2}{*}{$0.83 \pm 0.10  \pm 0.07$ }  &  &  \\
23$-$24
&   &   &   &  &  \\  \cline{2-4}
24$-$25
& \multirow{2}{*}{ $0.77 \pm 0.11  \pm 0.06$ }  & \multirow{2}{*}{$0.46 \pm 0.07  \pm 0.03$ }  & \multirow{2}{*}{$0.51 \pm 0.07  \pm 0.04$ }  &  &  \\  \cline{5-5}
25$-$26
&   &   &  & \multirow{5}{*}{$0.16 \pm 0.04  \pm 0.02$ } &  \\  \cline{2-4}
26$-$27
& \multirow{4}{*}{$0.44 \pm 0.06  \pm 0.04$ }  & \multirow{2}{*}{$0.41 \pm 0.07  \pm 0.03$ }  & \multirow{4}{*}{$0.23 \pm 0.04  \pm 0.02$ } &  &  \\
27$-$28
&   &   &  &  &  \\  \cline{3-3}
28$-$29
&   & \multirow{2}{*}{$0.32 \pm 0.06  \pm 0.03$ }  &  &  &  \\
29$-$30
 &   &   &  &  & \\\hline 
\end{tabular*}
 \caption{Double-differential cross-sections times dimuon branching fraction in different bins of \pt and $y$ for $\ThreeS$ (in~\pb). The first uncertainty is statistical and the second is systematic.}
 \label{tab:CS33}
 \end{small}
\end{sidewaystable}

%% file: Ratio1381.tex
 \begin{sidewaystable}[!ht]
 \centering
 \begin{small}
   \begin{tabular*}{0.99\textwidth}{|@{\hspace{1mm}}c@{\extracolsep{\fill}}ccccc@{\hspace{1mm}}|}
      \hline
      $\pt\left[\!\gevc\right]$  
      &  $2.0<y<2.5$
      &  $2.5<y<3.0$
      &  $3.0<y<3.5$
      &  $3.5<y<4.0$
      &  $4.0<y<4.5$  
      \\
      \hline 
0$-$1
& $1.231 \pm 0.081$& $1.200 \pm 0.071$& $1.257 \pm 0.081$& $1.346 \pm 0.109$& $1.578 \pm 0.188$\\ 
1$-$2
& $1.187 \pm 0.072$& $1.256 \pm 0.070$& $1.330 \pm 0.082$& $1.354 \pm 0.113$& $1.568 \pm 0.174$\\ 
2$-$3
& $1.256 \pm 0.075$& $1.241 \pm 0.069$& $1.325 \pm 0.080$& $1.484 \pm 0.114$& $1.575 \pm 0.203$\\ 
3$-$4
& $1.299 \pm 0.078$& $1.282 \pm 0.070$& $1.379 \pm 0.080$& $1.512 \pm 0.109$& $1.704 \pm 0.211$\\ 
4$-$5
& $1.345 \pm 0.079$& $1.334 \pm 0.074$& $1.420 \pm 0.082$& $1.555 \pm 0.116$& $1.919 \pm 0.235$\\ 
5$-$6
& $1.393 \pm 0.085$& $1.368 \pm 0.076$& $1.455 \pm 0.085$& $1.616 \pm 0.117$& $1.894 \pm 0.245$\\ 
6$-$7
& $1.484 \pm 0.087$& $1.408 \pm 0.078$& $1.540 \pm 0.089$& $1.654 \pm 0.124$& $1.993 \pm 0.258$\\ 
7$-$8
& $1.544 \pm 0.095$& $1.483 \pm 0.083$& $1.591 \pm 0.091$& $1.700 \pm 0.120$& $1.912 \pm 0.231$\\ 
8$-$9
& $1.566 \pm 0.096$& $1.504 \pm 0.086$& $1.649 \pm 0.096$& $1.709 \pm 0.133$& $2.171 \pm 0.279$\\ 
9$-$10
& $1.561 \pm 0.098$& $1.484 \pm 0.086$& $1.616 \pm 0.098$& $1.822 \pm 0.147$& $2.453 \pm 0.353$\\ 
10$-$11
& $1.633 \pm 0.112$& $1.637 \pm 0.097$& $1.723 \pm 0.106$& $1.899 \pm 0.160$& $2.592 \pm 0.343$\\ 
11$-$12
& $1.614 \pm 0.112$& $1.644 \pm 0.098$& $1.759 \pm 0.113$& $1.941 \pm 0.165$& $3.396 \pm 0.476$\\ 
12$-$13
& $1.702 \pm 0.122$& $1.648 \pm 0.102$& $1.752 \pm 0.115$& $1.959 \pm 0.179$& $2.712 \pm 0.421$
\\ \cline{6-6}
13$-$14 
& $1.691 \pm 0.130$& $1.540 \pm 0.099$& $1.777 \pm 0.123$& $2.026 \pm 0.194$
& \multirow{2}{*}{ $3.973 \pm 0.589$ } \\
14$-$15 
& $1.660 \pm 0.136$& $1.651 \pm 0.109$& $1.812 \pm 0.130$& $2.188 \pm 0.217$
& \\  \cline{6-6}
15$-$16
& $1.763 \pm 0.150$& $1.573 \pm 0.110$& $1.805 \pm 0.136$& $2.219 \pm 0.243$
& \multirow{5}{*}{$4.978 \pm 0.794$ } \\
16$-$17
& $2.018 \pm 0.174$& $1.790 \pm 0.132$& $1.865 \pm 0.151$& $2.168 \pm 0.260$
 &  \\
17$-$18
& $1.834 \pm 0.170$& $1.790 \pm 0.138$& $2.131 \pm 0.183$& $2.332 \pm 0.296$
 &  \\
18$-$19
& $1.529 \pm 0.150$& $1.984 \pm 0.166$& $1.877 \pm 0.180$& $2.679 \pm 0.376$
 &  \\
19$-$20
& $1.874 \pm 0.197$& $1.854 \pm 0.164$& $2.044 \pm 0.205$& $2.598 \pm 0.369$
 &  \\  \cline{2-6}
20$-$21
& \multirow{2}{*}{$1.853 \pm 0.176$ }  & \multirow{2}{*}{ $1.912 \pm 0.157$ }  & \multirow{2}{*}{ $2.088 \pm 0.196$ }  & \multirow{5}{*}{$2.865 \pm 0.369$ } & \\
21$-$22
&   &   &   &  &  \\  \cline{2-4}
22$-$23
& \multirow{2}{*}{$2.119 \pm 0.230$ }  & \multirow{2}{*}{$2.071 \pm 0.191$ }  & \multirow{2}{*}{$2.212 \pm 0.244$ }  &  &  \\
23$-$24
&   &   &   &  &  \\  \cline{2-4}
24$-$25
& \multirow{2}{*}{$1.664 \pm 0.210$ }  & \multirow{2}{*}{$1.863 \pm 0.216$ }  & \multirow{2}{*}{$2.652 \pm 0.335$ }  &  &  \\  \cline{5-5}
25$-$26
&   &   &  &  &  \\  \cline{2-4}
26$-$27
& \multirow{4}{*}{$2.340 \pm 0.274$ }  & \multirow{2}{*}{$2.059 \pm 0.262$}  & \multirow{4}{*}{$2.205 \pm 0.289$} &  &  \\
27$-$28
&   &   &  &  &  \\  \cline{3-3}
28$-$29
&   & \multirow{2}{*}{$2.096 \pm 0.326$ }  &  &  &  \\
29$-$30
 &   &   &  &  & \\\hline 
\end{tabular*}
\caption{The cross-section ratios between 13 TeV and 8 TeV in different bins of \pt and $y$ for \OneS.}
\label{tab:Ratio1381}
\end{small}
\end{sidewaystable}

%% file: Ratio1382.tex
 \begin{sidewaystable}[!ht]
 \centering
 \begin{small}
   \begin{tabular*}{0.99\textwidth}{|@{\hspace{1mm}}c@{\extracolsep{\fill}}ccccc@{\hspace{1mm}}|}
      \hline
      $\pt\left[\!\gevc\right]$  
      &  $2.0<y<2.5$
      &  $2.5<y<3.0$
      &  $3.0<y<3.5$
      &  $3.5<y<4.0$
      &  $4.0<y<4.5$  
      \\
      \hline 
0$-$1
& $1.177 \pm 0.114$& $1.106 \pm 0.086$& $1.177 \pm 0.100$& $1.318 \pm 0.128$& $1.194 \pm 0.185$\\ 
1$-$2
& $1.182 \pm 0.086$& $1.150 \pm 0.073$& $1.155 \pm 0.082$& $1.260 \pm 0.115$& $1.560 \pm 0.201$\\ 
2$-$3
& $1.187 \pm 0.082$& $1.105 \pm 0.068$& $1.267 \pm 0.085$& $1.329 \pm 0.112$& $1.454 \pm 0.203$\\ 
3$-$4
& $1.198 \pm 0.083$& $1.177 \pm 0.071$& $1.348 \pm 0.086$& $1.433 \pm 0.116$& $1.426 \pm 0.196$\\ 
4$-$5
& $1.251 \pm 0.085$& $1.243 \pm 0.076$& $1.390 \pm 0.087$& $1.460 \pm 0.121$& $1.864 \pm 0.252$\\ 
5$-$6
& $1.299 \pm 0.092$& $1.363 \pm 0.084$& $1.436 \pm 0.093$& $1.524 \pm 0.124$& $1.777 \pm 0.249$\\ 
6$-$7
& $1.472 \pm 0.106$& $1.330 \pm 0.084$& $1.433 \pm 0.093$& $1.636 \pm 0.138$& $1.847 \pm 0.268$\\ 
7$-$8
& $1.379 \pm 0.102$& $1.446 \pm 0.092$& $1.559 \pm 0.101$& $1.699 \pm 0.140$& $1.878 \pm 0.260$\\ 
8$-$9
& $1.390 \pm 0.108$& $1.473 \pm 0.097$& $1.605 \pm 0.106$& $1.689 \pm 0.152$& $1.978 \pm 0.308$\\ 
9$-$10
& $1.485 \pm 0.120$& $1.518 \pm 0.102$& $1.667 \pm 0.115$& $1.834 \pm 0.167$& $2.173 \pm 0.370$\\ 
10$-$11
& $1.470 \pm 0.130$& $1.553 \pm 0.110$& $1.589 \pm 0.115$& $1.913 \pm 0.184$& $2.899 \pm 0.473$\\ 
11$-$12
& $1.526 \pm 0.142$& $1.538 \pm 0.111$& $1.694 \pm 0.132$& $1.775 \pm 0.185$& $2.848 \pm 0.545$\\ 
12$-$13
& $1.709 \pm 0.162$& $1.707 \pm 0.126$& $1.627 \pm 0.132$& $1.845 \pm 0.203$& $3.014 \pm 0.675$
\\ \cline{6-6}
13$-$14 
& $1.612 \pm 0.173$& $1.662 \pm 0.133$& $1.625 \pm 0.144$& $1.974 \pm 0.233$
& \multirow{2}{*}{  $4.958 \pm 0.959$ } \\
14$-$15 
& $1.607 \pm 0.186$& $1.671 \pm 0.143$& $1.986 \pm 0.175$& $1.995 \pm 0.257$
& \\  \cline{6-6}
15$-$16
& $1.681 \pm 0.201$& $1.666 \pm 0.147$& $1.964 \pm 0.192$& $1.863 \pm 0.268$
& \multirow{5}{*}{$4.896 \pm 1.034$ } \\
16$-$17
& $1.541 \pm 0.196$& $1.639 \pm 0.161$& $1.897 \pm 0.201$& $1.754 \pm 0.294$
 &  \\
17$-$18
& $1.819 \pm 0.253$& $1.442 \pm 0.155$& $1.894 \pm 0.221$& $2.348 \pm 0.402$
 &  \\
18$-$19
& $1.768 \pm 0.246$& $1.918 \pm 0.210$& $1.713 \pm 0.215$& $2.098 \pm 0.414$
 &  \\
19$-$20
& $2.158 \pm 0.307$& $2.073 \pm 0.243$& $2.110 \pm 0.317$& $2.507 \pm 0.494$
 &  \\  \cline{2-6}
20$-$21
& \multirow{2}{*}{$1.529 \pm 0.208$ }  & \multirow{2}{*}{ $1.749 \pm 0.184$ }  & \multirow{2}{*}{ $2.396 \pm 0.284$ }  & \multirow{5}{*}{$3.504 \pm 0.536$ } & \\
21$-$22
&   &   &   &  &  \\  \cline{2-4}
22$-$23
& \multirow{2}{*}{$2.323 \pm 0.349$ }  & \multirow{2}{*}{$1.720 \pm 0.229$ }  & \multirow{2}{*}{$1.914 \pm 0.304$ }  &  &  \\
23$-$24
&   &   &   &  &  \\  \cline{2-4}
24$-$25
& \multirow{2}{*}{$1.353 \pm 0.244$ }  & \multirow{2}{*}{$1.923 \pm 0.288$ }  & \multirow{2}{*}{$2.269 \pm 0.387$ }  &  &  \\  \cline{5-5}
25$-$26
&   &   &  &  &  \\  \cline{2-4}
26$-$27
& \multirow{4}{*}{$1.710 \pm 0.312$ }  & \multirow{2}{*}{$2.088 \pm 0.364$}  & \multirow{4}{*}{$2.294 \pm 0.420$} &  &  \\
27$-$28
&   &   &  &  &  \\  \cline{3-3}
28$-$29
&   & \multirow{2}{*}{ $2.538 \pm 0.517$ }  &  &  &  \\
29$-$30
 &   &   &  &  & \\\hline 
\end{tabular*}
\caption{The cross-section ratios between 13 TeV and 8 TeV in different bins of \pt and $y$ for \TwoS.}
\label{tab:Ratio1382}
\end{small}
\end{sidewaystable}

%% file: Ratio1383.tex
 \begin{sidewaystable}[!ht]
 \centering
 \begin{small}
   \begin{tabular*}{0.99\textwidth}{|@{\hspace{1mm}}c@{\extracolsep{\fill}}ccccc@{\hspace{1mm}}|}
      \hline
      $\pt\left[\!\gevc\right]$  
      &  $2.0<y<2.5$
      &  $2.5<y<3.0$
      &  $3.0<y<3.5$
      &  $3.5<y<4.0$
      &  $4.0<y<4.5$  
      \\
      \hline 
0$-$1
& $1.117 \pm 0.167$& $1.145 \pm 0.126$& $1.362 \pm 0.174$& $1.265 \pm 0.173$& $1.681 \pm 0.328$\\ 
1$-$2
& $1.372 \pm 0.135$& $1.143 \pm 0.095$& $1.138 \pm 0.109$& $1.364 \pm 0.152$& $1.566 \pm 0.251$\\ 
2$-$3
& $1.294 \pm 0.119$& $1.275 \pm 0.096$& $1.389 \pm 0.116$& $1.397 \pm 0.139$& $1.399 \pm 0.223$\\ 
3$-$4
& $1.193 \pm 0.107$& $1.349 \pm 0.099$& $1.386 \pm 0.108$& $1.409 \pm 0.136$& $1.702 \pm 0.264$\\ 
4$-$5
& $1.443 \pm 0.124$& $1.403 \pm 0.103$& $1.299 \pm 0.101$& $1.475 \pm 0.141$& $1.633 \pm 0.260$\\ 
5$-$6
& $1.601 \pm 0.139$& $1.477 \pm 0.109$& $1.539 \pm 0.118$& $1.514 \pm 0.148$& $1.896 \pm 0.308$\\ 
6$-$7
& $1.495 \pm 0.135$& $1.524 \pm 0.115$& $1.539 \pm 0.119$& $1.676 \pm 0.165$& $2.131 \pm 0.351$\\ 
7$-$8
& $1.618 \pm 0.149$& $1.376 \pm 0.108$& $1.643 \pm 0.125$& $1.680 \pm 0.160$& $2.198 \pm 0.355$\\ 
8$-$9
& $1.593 \pm 0.152$& $1.438 \pm 0.116$& $1.704 \pm 0.132$& $1.565 \pm 0.169$& $2.494 \pm 0.447$\\ 
9$-$10
& $1.724 \pm 0.170$& $1.607 \pm 0.130$& $1.689 \pm 0.138$& $1.609 \pm 0.176$& $2.538 \pm 0.533$\\ 
10$-$11
& $1.797 \pm 0.199$& $1.431 \pm 0.125$& $1.566 \pm 0.136$& $1.645 \pm 0.186$& $2.253 \pm 0.474$\\ 
11$-$12
& $1.755 \pm 0.210$& $1.687 \pm 0.143$& $1.808 \pm 0.166$& $1.610 \pm 0.196$& $2.930 \pm 0.689$\\ 
12$-$13
& $1.451 \pm 0.184$& $1.639 \pm 0.146$& $1.870 \pm 0.172$& $1.933 \pm 0.239$& $3.684 \pm 0.985$
\\ \cline{6-6}
13$-$14 
& $1.177 \pm 0.174$& $1.727 \pm 0.158$& $1.852 \pm 0.189$& $2.167 \pm 0.291$
& \multirow{2}{*}{ $3.385 \pm 0.851$ } \\
14$-$15 
& $1.478 \pm 0.218$& $1.580 \pm 0.163$& $1.574 \pm 0.175$& $2.100 \pm 0.311$
& \\  \cline{6-6}
15$-$16
& $1.680 \pm 0.245$& $1.848 \pm 0.199$& $1.980 \pm 0.231$& $1.941 \pm 0.328$
& \multirow{5}{*}{$4.640 \pm 1.254$ } \\
16$-$17
& $2.348 \pm 0.337$& $1.587 \pm 0.189$& $1.888 \pm 0.232$& $2.167 \pm 0.419$
 &  \\
17$-$18
& $1.809 \pm 0.315$& $1.824 \pm 0.227$& $1.827 \pm 0.262$& $1.913 \pm 0.377$
 &  \\
18$-$19
& $1.958 \pm 0.319$& $2.109 \pm 0.274$& $1.937 \pm 0.304$& $1.977 \pm 0.460$
 &  \\
19$-$20
& $1.987 \pm 0.357$& $1.696 \pm 0.244$& $2.075 \pm 0.369$& $2.116 \pm 0.536$
 &  \\  \cline{2-6}
20$-$21
& \multirow{2}{*}{$1.532 \pm 0.254$ }  & \multirow{2}{*}{ $1.846 \pm 0.234$ }  & \multirow{2}{*}{ $2.536 \pm 0.348$ }  & \multirow{5}{*}{$3.255 \pm 0.582$ } & \\
21$-$22
&   &   &   &  &  \\  \cline{2-4}
22$-$23
& \multirow{2}{*}{$1.744 \pm 0.346$ }  & \multirow{2}{*}{$2.294 \pm 0.313$ }  & \multirow{2}{*}{$2.580 \pm 0.412$ }  &  &  \\
23$-$24
&   &   &   &  &  \\  \cline{2-4}
24$-$25
& \multirow{2}{*}{$2.257 \pm 0.466$ }  & \multirow{2}{*}{ $1.510 \pm 0.285$ }  & \multirow{2}{*}{$2.408 \pm 0.461$ }  &  &  \\  \cline{5-5}
25$-$26
&   &   &  &  &  \\  \cline{2-4}
26$-$27
& \multirow{4}{*}{$1.700 \pm 0.316$ }  & \multirow{2}{*}{$2.215 \pm 0.459$}  & \multirow{4}{*}{$2.339 \pm 0.495$} &  &  \\
27$-$28
&   &   &  &  &  \\  \cline{3-3}
28$-$29
&   & \multirow{2}{*}{$2.629 \pm 0.654$ }  &  &  &  \\
29$-$30
 &   &   &  &  & \\\hline 
\end{tabular*}
\caption{The cross-section ratios between 13 TeV and 8 TeV in different bins of \pt and $y$ for \ThreeS.}
\label{tab:Ratio1383}
\end{small}
\end{sidewaystable}

%% file: acknowledgements.tex
\section*{Acknowledgements}
%
%
\noindent We thank Kuang-Ta Chao, Yu Feng, Yan-Qing Ma, Hua-Sheng Shao
and Jian-Xiong Wang for frequent and interesting discussions
on the production of $\PUpsilon$ mesons.
We express our gratitude to our colleagues in the CERN
accelerator departments for the excellent performance of the LHC. We
thank the technical and administrative staff at the LHCb
institutes. We acknowledge support from CERN and from the national
agencies: CAPES, CNPq, FAPERJ and FINEP (Brazil); MOST and NSFC
(China); CNRS/IN2P3 (France); BMBF, DFG and MPG (Germany); INFN
(Italy); NWO (The Netherlands); MNiSW and NCN (Poland); MEN/IFA
(Romania); MinES and FASO (Russia); MinECo (Spain); SNSF and SER
(Switzerland); NASU (Ukraine); STFC (United Kingdom); NSF (USA).  We
acknowledge the computing resources that are provided by CERN, IN2P3
(France), KIT and DESY (Germany), INFN (Italy), SURF (The
Netherlands), PIC (Spain), GridPP (United Kingdom), RRCKI and Yandex
LLC (Russia), CSCS (Switzerland), IFIN-HH (Romania), CBPF (Brazil),
PL-GRID (Poland) and OSC (USA). We are indebted to the communities
behind the multiple open-source software packages on which we depend.
Individual groups or members have received support from AvH Foundation
(Germany), EPLANET, Marie Sk\l{}odowska-Curie Actions and ERC
(European Union), ANR, Labex P2IO and OCEVU, and R\'{e}gion
Auvergne-Rh\^{o}ne-Alpes (France), Key Research Program of Frontier
Sciences of CAS, CAS PIFI, and the Thousand Talents Program (China),
RFBR, RSF and Yandex LLC (Russia), GVA, XuntaGal and GENCAT (Spain),
Herchel Smith Fund, the Royal Society, the English-Speaking Union and
the Leverhulme Trust (United Kingdom).

%% file: LHCb_Authorship_flat_22-Jan-2018.tex
\centerline{\large\bf LHCb collaboration}
\begin{flushleft}
\small
R.~Aaij$^{43}$,
B.~Adeva$^{39}$,
M.~Adinolfi$^{48}$,
Z.~Ajaltouni$^{5}$,
S.~Akar$^{59}$,
J.~Albrecht$^{10}$,
F.~Alessio$^{40}$,
M.~Alexander$^{53}$,
A.~Alfonso~Albero$^{38}$,
S.~Ali$^{43}$
% ...

{\footnotesize \it
% ...
$ ^{5}$Clermont Universit{\'e}, Universit{\'e} Blaise Pascal, CNRS/IN2P3, LPC, Clermont-Ferrand, France\\
% ...
$ ^{10}$Fakult{\"a}t Physik, Technische Universit{\"a}t Dortmund, Dortmund, Germany\\
% ...
$ ^{38}$ICCUB, Universitat de Barcelona, Barcelona, Spain\\
$ ^{39}$Instituto Galego de F{\'\i}sica de Altas Enerx{\'\i}as (IGFAE), Universidade de Santiago de Compostela, Santiago de Compostela, Spain\\
$ ^{40}$European Organization for Nuclear Research (CERN), Geneva, Switzerland\\
% ...
$ ^{43}$Nikhef National Institute for Subatomic Physics, Amsterdam, The Netherlands\\
% ...
$ ^{48}$H.H. Wills Physics Laboratory, University of Bristol, Bristol, United Kingdom\\
% ...
$ ^{53}$School of Physics and Astronomy, University of Glasgow, Glasgow, United Kingdom\\
% ...
$ ^{59}$University of Cincinnati, Cincinnati, OH, United States\\
% ...
}
\end{flushleft}